\pdfoutput=1
\documentclass[prb,twocolumn,showpacs,preprintnumbers,superscriptaddress,amsmath,amssymb]{revtex4}
\usepackage{graphicx}
\usepackage{float}
\usepackage{dcolumn}
\usepackage{color}
\usepackage[dvipsnames]{xcolor} % 声明颜色包
\usepackage{latexsym,bm}
\usepackage[normalem]{ulem}
\usepackage{multirow}
\usepackage{appendix}
\usepackage{amsmath}
\usepackage{amsfonts}
\usepackage{booktabs}
\usepackage{subcaption}
\usepackage{tabularx}

\usepackage{threeparttable} % to use table notes

\newcommand{\mc}[1]{\textcolor{black}{{#1}}}
\newcommand{\SL}[1]{\textcolor{black}{{#1}}}

\newcommand{\tx}[1]{\text{#1}}

\newcommand{\TKK}[2]{\ensuremath{\text{TKK}^{\text{#1}}_{#2}}}

\begin{document}

\hyphenpenalty=5000
\tolerance=1000

\title{Truncated Non-Local Kinetic Energy Density Functionals for \mc{Simple Metals and Silicon}}
% \title{Truncated Non-Local Kinetic Energy Density Functionals for Metals and Semiconductors}
%\title{Unified Approach for Designing Kinetic Energy Density Functional for Orbital-Free Density Functional Theory}
%\title{Truncated Non-Local Kinetic Energy Density Functional for Orbital-Free Density Functional Theory}

\author{Liang Sun}
\affiliation{HEDPS, CAPT, School of Physics and College of Engineering, Peking University, Beijing 100871, P. R. China}
\author{Yuanbo Li}
\affiliation{HEDPS, CAPT, School of Physics and College of Engineering, Peking University, Beijing 100871, P. R. China}
\author{Mohan Chen}
\email{mohanchen@pku.edu.cn}
\affiliation{HEDPS, CAPT, School of Physics and College of Engineering, Peking University, Beijing 100871, P. R. China}
\affiliation{AI for Science Institute, Beijing 100080, P. R. China}
\date{\today}
%%%%%%%%%%%%%%%%%%%%%%%%%%%%%%%%%%%%%%%%%%%%%%%%%%%%%%%%%%%%%%%%%%%%%%%%%%%%%%%%%%%%%%%%%%%%%%%%%%%%%%%%%%%%%%%%%%%%%%%%%%%%%%%%%%%
%%%%%     Title
%%%%%%%%%%%%%%%%%%%%%%%%%%%%%%%%%%%%%%%%%%%%%%%%%%%%%%%%%%%%%%%%%%%%%%%%%%%%%%%%%%%%%%%%%%%%%%%%%%%%%%%%%%%%%%%%%%%%%%%%%%%%%%%%%%%

\begin{abstract}
Adopting an accurate kinetic energy density functional (KEDF) to characterize the noninteracting kinetic energy within the framework of orbital-free density functional theory (OFDFT) is challenging. 
We propose a new form of the non-local KEDF with a real-space truncation cutoff that satisfies the uniform electron gas limit and design KEDFs for \mc{simple metals and silicon}.
The new KEDFs are obtained by minimizing a residual function, which contains the differences in the total energy and charge density of several representative systems with respect to the Kohn-Sham DFT results.
By systematically testing different cutoffs of the new KEDFs, we find that the cutoff plays a crucial role in determining the properties of metallic Al and semiconductor Si systems.
We conclude that the new KEDF with a sufficiently long cutoff performs even better than some representative non-local KEDFs in some aspects, which sheds new light on optimizing the KEDFs in OFDFT to achieve better accuracy.
\end{abstract}
%\pacs{}
\maketitle

\section{Introduction}
Kohn-Sham density functional theory (KSDFT) is one of the most widely used {\it ab initial} methods.~\cite{64PR-Hohenberg,65PR-Kohn}
However, since the traditional KSDFT method introduces orthogonal one-electron orbitals, solving the Kohn-Sham equation typically scales as $O(N^3)$ with $N$ being the atom number, which is unfavorable for large-size calculations or long-time molecular dynamics simulations.
Orbital-free DFT (OFDFT)~\cite{02Carter,18JMR-Witt} is an alternative choice to improve the efficiency of DFT by calculating the non-interacting electron kinetic energy $T_s$ via the kinetic energy density functional (KEDF) instead of the one-electron Kohn-Sham orbitals.
OFDFT has been successfully applied to a variety of scientific problems such as alloys,~\cite{14AM-Shin, 17MSMSE-Zhuang, 18PRM-Zhuang, 18JMR-Witt, 21JPCA-Witt} liquid metals,~\cite{13MP-Chen} quantum dots,~\cite{19B-Mi-qdot, 20B-Xu-nonlocal} and warm dense matter.\cite{20JPCM-QianruiLiu-wdm, 20Kang-semilocal} Recently, the time-dependent OFDFT has been proposed to study the stopping power of electrons in warm dense matter,~\cite{18L-Ding-oftddft,18B-White-oftddft} the localized surface plasmon resonances in nanorods,~\cite{20JPC-Xiang-oftddft} and the optical spectra of metallic and semiconductor clusters.~\cite{21B-Jiang-oftddft}
Since the magnitude of electron kinetic energy ($T_s$) is comparable to the total energy in condensed matter and molecular systems, the accuracy of OFDFT is sensitive to the approximated forms of the KEDF. In this regard,
proposing an accurate and efficient KEDF within the framework of OFDFT has been a challenging topic in this community for decades.

In the past few decades, continuous efforts have been devoted to the development of KEDFs. As a result, various forms of KEDF were proposed.
A typical category of KEDFs includes the local and semi-local components, which can be efficiently evaluated. For instance, Constantin et al. demonstrated the importance of adopting the Laplacian of charge density in the construction of KEDFs and proposed a series of new semi-local KEDFs~\cite{18JPCL-Constantin-semilocal, 19JCTC-Constantin-semilocal}. 
Luo et al. generalized the LKT KEDF~\cite{18B-Luo-semilocal} to finite temperatures~\cite{20B-Luo-semilocal} and applied it to the warm dense hydrogen.~\cite{20Kang-semilocal}
% Luo et al. generalised the LKT KEDF~\cite{18B-Luo-semilocal} to finite temperatures~\cite{20B-Luo-semilocal} suitable for studying the warm dense hydrogen.~\cite{20Kang-semilocal} (the performance of the KEDF is not very good)
%
The next category of KEDFs is the non-local form, which suggests that the kinetic density at each real-space point depends on the non-local charge density.
For condensed matter systems, the non-local KEDFs are generally more accurate than the semi-local ones, such as the Wang-Teter (WT)~\cite{92B-Wang-nonlocal}, the Smargiassi-Madden (SM)~\cite{94B-Smargiassi-nonlocal}, and the Wang-Govind-Carter (WGC)~\cite{99B-Wang-nonlocal} KEDFs for metals and the Huang-Carter (HC) KEDF\cite{10B-Huang-nonlocal} for semiconductors.
While most KEDFs were constructed based on the Lindhard response function, another category of KEDFs was introduced by imposing more restrictions or using more parameters.
For example, the enhanced von Weizsäcker WGC KEDF,~\cite{14JCP-Shin-nonlocal} the KGAP KEDF based on the jellium-with-gap model,~\cite{18B-Constantin-nonlocal} the KEDFs in the form of functional integrals,~\cite{18JCP-Mi-nonlocal, 19B-Xu-nonlocal}. In particular, the revised HC KEDF~\cite{21B-Shao-nonlocal} was proposed to achieve higher precision for the surface of semiconductors.

While most of the non-local KEDFs implement a non-local kinetic energy kernel, a fundamental yet important issue regarding how the long- and short-ranged parts of the kinetic energy kernel influence the accuracy of KEDFs is still unclear.
In this regard, a truncated KEDF kernel (TKK) with a chosen real-space cutoff could provide further information for this issue.
Recently, a truncated WT kernel was proposed~\cite{16JCTC-Chen} to enable efficient calculations of 1,024,000 lithium (Li) atoms with the OFDFT method. 
%In particular, a small-box-based Fast Fourier transform technique was used.
%
The truncated WT kernel is composed of eight Spherical Bessel functions and yields reasonable results for Li systems.
\mc{Note that the Spherical Bessel functions have been used as localized basis sets in density functional theory calculations.~\cite{97CPC-Haynes-bessel,10JPCM-Chen-bessel, 11JPCM-Chen-bessel, 16CMS-Li-bessel}}
Similarly, Kumar et al. proposed a non-local KEDF 
whose kinetic energy kernel consists of six Gaussian functions and found improved performances for a series of one-dimensional systems.~\cite{22JCP-Kumar-Gaussiankernel}
In this regard, constructing a TKK that owns sufficient accuracy within the framework of OFDFT has been demonstrated to be feasible; nevertheless, an important remaining issue is to reveal the influences of the long- and short-ranged parts of KEDF on a selection of target systems.

%-------------
% Figure 1
%-------------
\begin{figure*}[htbp]
	\centering
	\includegraphics[width=1.0\textwidth]{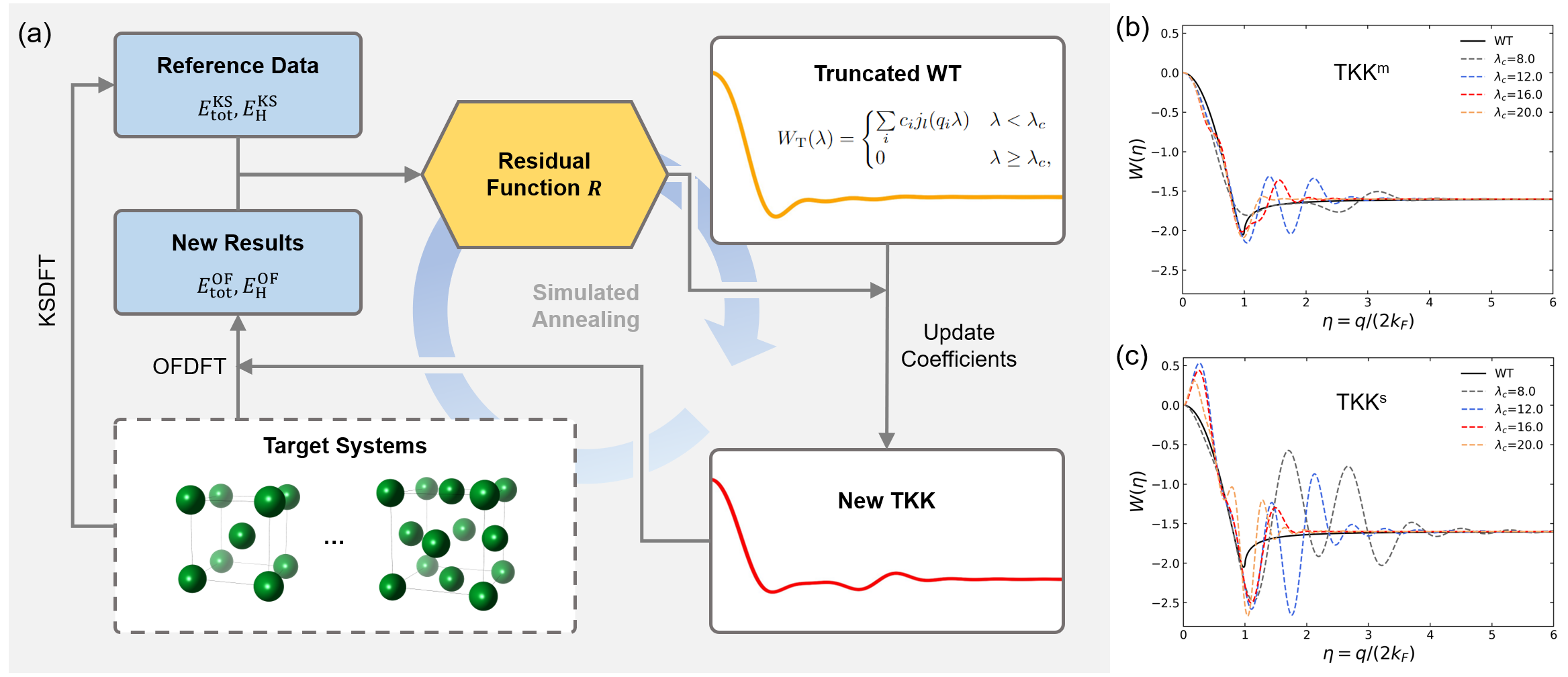}\\
	\caption{
	(a) Workflow of the simulated annealing method to optimize the truncated KEDF kernel (TKK) in the framework of OFDFT.
    (b) The fitted TKKs in reciprocal space for metallic systems (labeled as \TKK{m}{}) with the target systems being Al systems. The Wang-Teter (WT) KEDF kernel is plotted for comparison. The real-space cutoff $\lambda_{c}$ is chosen with different values ranging from 8 to 20.
    (c) The fitted TKKs in reciprocal space for semiconductor systems (labeled as \TKK{s}{}) with the target systems being Si systems.
	}\label{fig:workflow}
\end{figure*}

In this work, we construct two groups of TKK, one for metals (labeled as TKK$^{\mathrm{m}}_{\lambda_c}$ with $\lambda_c$ being the real-space cutoff) and the other for semiconductors (labeled as TKK$^{\mathrm{s}}_{\lambda_c}$) because the asymptotic behaviors of KEDFs for metals and semiconductors are different.~\cite{10B-Huang-nonlocal}
In particular, \mc{as a first step to find an optimal KEDF for metals and another one for semiconductors,} we respectively choose Al and Si to validate the two groups of TKKs.
For each group, a few TKKs are represented by a set of spherical Bessel functions and generated~\cite{16JCTC-Chen} with different radius cutoffs. The coefficients of spherical Bessel functions are optimized with the simulated annealing method.~\cite{53JCP-Metropolis, 83Science-Kirkpatrick}
We systematically test these kernels for a variety of Al and Si systems.
In general, the accuracy of TKKs increases with a larger cutoff.
In particular, we find it crucial to consider the interactions between an atom and its nearest neighbors, as well as the next nearest neighbors in a TKK; otherwise, the stacking fault energies and surface energies of Al, as well as the vacancy formation energies and surface energies of Si are qualitatively incorrect.
Additionally, the TKK$^{\mathrm{m}}_{16}$ kernel works well for Li and Mg bulk systems, demonstrating its transferability.
The computational efficiency of TKKs is similar to the WT and WGC KEDFs, and higher than the HC KEDF.

The rest of this paper is organized as follows.
In Section II, we introduce the method to optimize the truncated WT kernel. In Section III, we list the numerical details of KSDFT and OFDFT calculations.
In Section IV, we analyze the performances of the new KEDF kernel and discuss the results.
%test performances of the optimized TKKs for a variety of systems, including the bulk properties of Al, Li, Mg, $\beta''\tx{-Al}_3\tx{Mg}$, $\tx{L1}_2\ \tx{Mg}_3\tx{Al}$ and Si, the surface energies and vacancy formation energies of fcc Al and CD Si, and the stacking fault energies of fcc Al.
%We analyze the performance of TKKs, which are related to the influence of long-range part in KEDF.
Finally, the conclusions are drawn in Section V.

\section{Methods}
\subsection{Kinetic Energy Density Functional Kernel}

The WT KEDF~\cite{92B-Wang-nonlocal} is derived from the Lindhard response function and takes the form of
\begin{equation}
\begin{aligned}
    T_{\tx{WT}}[\rho(&{\bf{r})}] = C_{\tx{TF}} \int {\rho^{5/3} ({\bf{r}}) \,\tx{d}{\bf{r}}} + \frac{1}{8} \int {\frac{{\left| {\nabla \rho ({\bf{r}})} \right|}^2}{\rho \,({\bf{r}})} \,\tx{d}{\bf{r}}} \\
    &+ C_{\tx{TF}}\iint{\rho ^{\alpha}({\bf{r}}){W}({\bf{r}} - {\bf{r'}}){\rho ^{\beta}}({\bf{r'}})\,{\tx{d}}{{\bf{r}}}\,{\tx{d}}{{\bf{r'}}}},
\end{aligned}
\end{equation}
where $C_{\tx{TF}} = \frac{3}{10}(3\pi^2)^{2/3}$. The parameters $\alpha$ and $\beta$ are typically set to $5/6$.
The first term is the Thomas-Fermi (TF) KEDF,~\cite{27TANL-Fermi-local}, which is a local functional exact for the uniform electron gas.
The second term is the semi-local von Weizs$\mathrm{\Ddot{a}}$cker (vW) KEDF,~\cite{35-vW-semilocal} which is a rigorous lower bound to the $T_s$.
The last term is a non-local form of KEDF derived from
the Lindhard response function with
$W(\bf{r} - \bf{r'})$ being the non-local KEDF kernel. Furthermore, \mc{the kernel can be analytically written in the reciprocal space as~\cite{92B-Wang-nonlocal,99B-Wang-nonlocal}}
\begin{equation}
    W(\eta) = \frac{5G(\eta)}{9\alpha \beta \rho_0^{\alpha+\beta-5/3}},
\end{equation}
where
\begin{equation}
    G(\eta) = {{\left( {\frac{1}{2} + \frac{1-\eta^2}{4\eta}\ln \left| {\frac{1 + \eta}{1 - \eta}} \right|} \right)}^{ - 1}} - 3\eta^2 - 1.
\end{equation}
Here $\eta = \frac{k}{2k_{\tx{F}}}$ is a dimensionless reciprocal space vector, while $k_{\tx{F}} = (3\pi^2\rho_0)^{1/3}$ is the Fermi wave vector with $\rho_0$ being the average charge density.
The truncated kinetic kernel is expressed as a linear combination of Spherical Bessel functions and takes the form of
\begin{equation}
    W_{\tx{T}}(\lambda ) =
    \begin{cases}
    \sum\limits_i {{c_i}{j_l}(q_i\lambda )} & \lambda < {\lambda _c}\\
    0                                     & \lambda \ge {\lambda _c},
    \end{cases}
    \label{eq.tkk}
\end{equation}
where ${j_l}(q_i\lambda)$ is a Spherical Bessel function and $c_{i}$ is the coefficient. The parameter $q_i$ satisfies ${j_l}(q_i\lambda_c) = 0$. Here the real-space cutoff is $\lambda_c = 2k_{\tx{F}} |{\bf{r}} - {\bf{r'}}|$. \mc{The $l$ parameter is set to 0, which is the same as in Ref.~\onlinecite{16JCTC-Chen}.}
%truncated WT kernel with $l = 0$ and $\lambda_c = 8.0$ was proven to perform well in Li systems.

\subsection{Residual Function}
With the aim of obtaining a more accurate KEDF kernel, we propose to optimize the coefficients of the Spherical Bessel functions in Eq.~\ref{eq.tkk} for a selected set of representative systems. In this regard, we define a residual function as
\mc{
\begin{equation}
    R = {\left| {\Delta {E_{{\tx{tot}}}}} \right|}  + \mu {\left| {\Delta {E_{\tx{H}}}} \right|}  +\nu |J| + \xi |G|,
    \label{eq.residual}
\end{equation}
where $\mu$, $\nu$, and $\xi$ are the coefficients.
}
Here the first term denotes the absolute total energy difference of a target system as calculated by the OFDFT and KSDFT methods, the formula is as follows
\begin{equation}
{\left| {\Delta {E_{\tx{tot}}}} \right|}  = \frac{1}{N}\sum\limits_{j = 1}^N {\left| {E^{\tx{OF}}_{\tx{tot},j} - E^{{\tx{KS}}}_{\tx{tot},j}} \right|}/n_j,
\end{equation}
where $E^{\tx{OF}}_{\tx{tot},j}$ and $E^{{\tx{KS}}}_{\tx{tot},j}$ are the total energies of the $j$th system as computed by the OFDFT utilizing the truncated kinetic kernels and the KSDFT method, respectively. $N$ is the number of selected reference systems and $n_j$ is the number of atoms in the $j$th system.

In order to minimize the charge density difference, we add a second term to minimize the absolute energy difference of the Hartree energy term, which takes the form of
\begin{equation}
{\left| {\Delta {E_{\tx{H}}}} \right|}  = 
\frac{1}{N}\sum\limits_{j = 1}^N {\left| {E_{\tx{H},j}^{\tx{OF}} - E_{{\tx{H},j}}^{\tx{KS}}} \right|}/n_j,
\end{equation}
where $E_{\tx{H},j}^{\tx{OF}}$ and $E_{{\tx{H},j}}^{\tx{KS}}$ represent the Hartree energies from the OFDFT and KSDFT calculations, respectively; both Hartree energies can be computed from $\frac{1}{2}\int\int \frac{\rho(\mathbf{r})\rho(\mathbf{r^{\prime}})}{|\mathbf{r}-\mathbf{r^{\prime}}|} d\mathbf{r}d\mathbf{r^{\prime}}$
with $\rho(\mathbf{r})$ being the charge density.

In fact, in a uniform electron gas, \mc{the electron density remains a constant and the TF model is exact to describe the kinetic energy of electrons. In this case, the nonlocal term is expected to disappear, and the TKK KEDF should be equivalent to the TF model. This implies that the integration of the nonlocal TKK should yield zero.}
In this regard, we impose a constraint in the third term $J$ \mc{to satisfy the limit of uniform electron gas}, which can be written as
\begin{equation}
J = \sum\limits_{i = 1}^{N_i} {{c_i}\int\limits_0^{\lambda _c} {{\lambda ^2}j_0(q_i\lambda ) \,{\tx{d}}\lambda}}.
\end{equation}
The term $J$ is the integral of a TKK in real space with $N_i$ being the number of Spherical Bessel functions. For a TKK that minimizes the residual function, the $J$ term is supposed to be zero to satisfy the abovementioned constraint.

Last, the fourth term, $G$, is included as a penalty term to reduce the oscillation behaviors of the truncated kinetic kernels and takes the form of
\begin{equation}
G=\sum\limits_{i = 1}^{N_i} {{c_i}\int\limits_0^{{\eta _c}} {{\eta ^4}\widehat F[ {j_0}({q_i}\lambda)] \,{\tx{d}}\eta}}.
\end{equation}
Here, $\widehat F$ donates the Fourier transform.
\mc{The last term is added because we find the resulting TKK kernel exhibits oscillating behavior in real space. Typically, the oscillating behavior can be effectively reduced in optimizing the shape of the function in reciprocal space. Therefore, we multiply the Fourier transform of the kernel function in reciprocal space by a $\eta^4$ term to reduce the oscillations especially when the dimensionless reciprocal space vector $\eta$ is large.}
\mc{In practice, we set the number of Spherical Bessel functions to be $N_i=8$ and the remaining parameters in Eq.~\ref{eq.residual}, i.e., $\mu$, $\nu$, and $\xi$, are respectively set to 3, $1/10$, and $1/20$, so that the above four terms account for similar proportions of the residual.}

\subsection{Simulated Annealing Method}
As illustrated in Fig.~\ref{fig:workflow}(a),
we adopt the simulated annealing method to minimize the residual function $R$ defined in Eq.~\ref{eq.residual}.
The workflow contains three steps. 
First, the KSDFT calculations are performed to yield the total energy $E^{\tx{KS}}_{\tx{tot}}$ and the Hartree energy $E^{\tx{KS}}_{\tx{H}}$ of selected target systems as reference data.
Second, starting from the truncated WT kernel, the coefficients $\{c_{i}\}$ of Spherical Bessel functions are updated to obtain a new TKK $W_{\tx{T}}(\lambda)$. We then use this new TKK and perform OFDFT calculations for target systems to obtain the residual function $R$ and the change of residual $\Delta R$.
Third, we use the Metropolis algorithm to update the coefficients $\{c_{i}\}$, which means that the previously updated coefficient will be accepted if $\Delta R \leq 0$; or if $\Delta R > 0$, the updated coefficients will be accepted with the probability $p={e^{ - \Delta R/T}}$, \mc{where $T$ is the artificial temperature and reduces gradually during the optimization.}

%-------------
% Figure 2
%-------------
\begin{figure*}[htbp]
    \centering

    \begin{subfigure}{0.49\textwidth}
    \centering
    \includegraphics[width=0.95\linewidth]{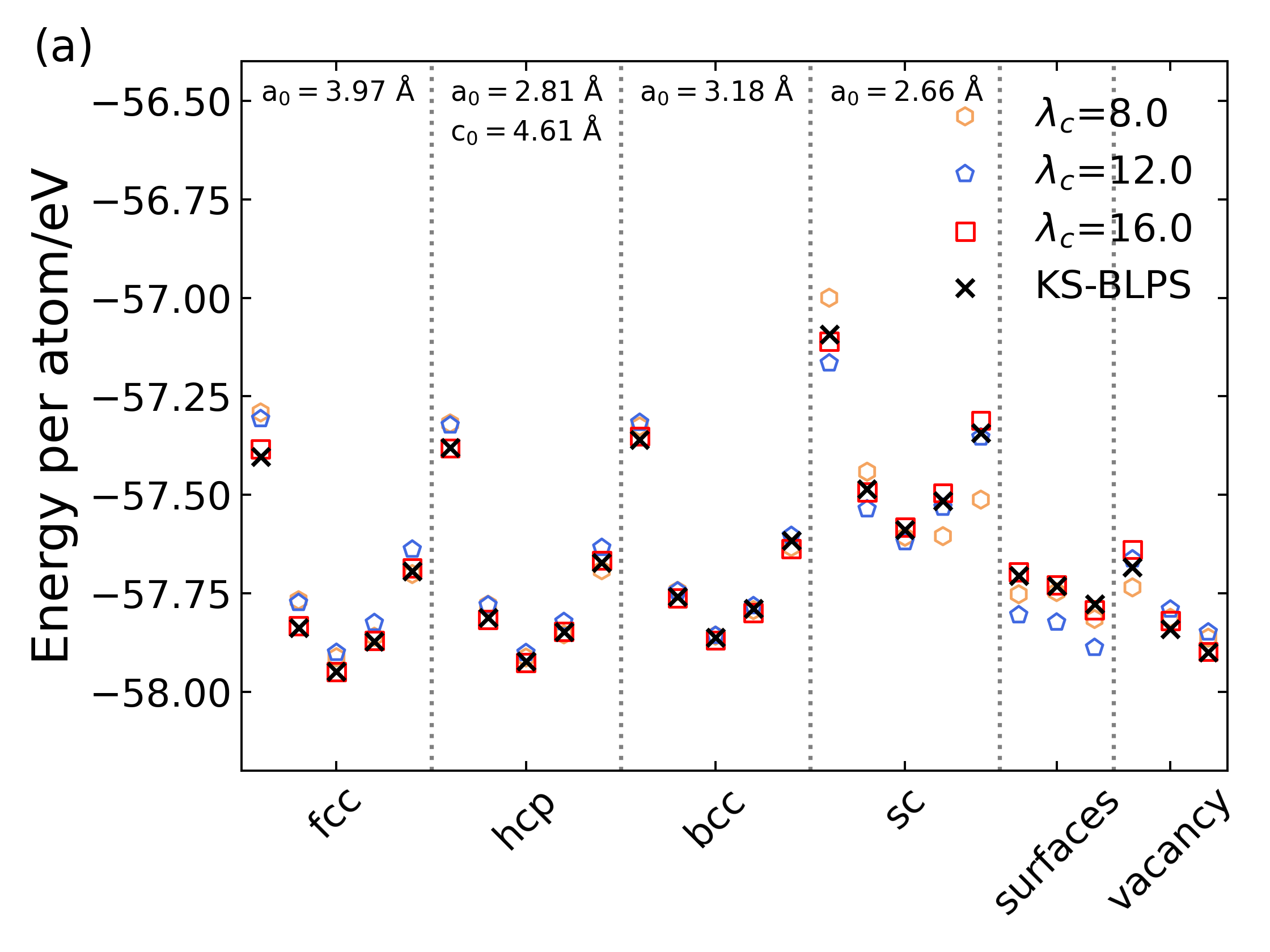}
    % \caption{}
    \label{fig:Fitting_Ala}
    \end{subfigure}
    \begin{subfigure}{0.49\textwidth}
    \centering
    \includegraphics[width=0.95\linewidth]{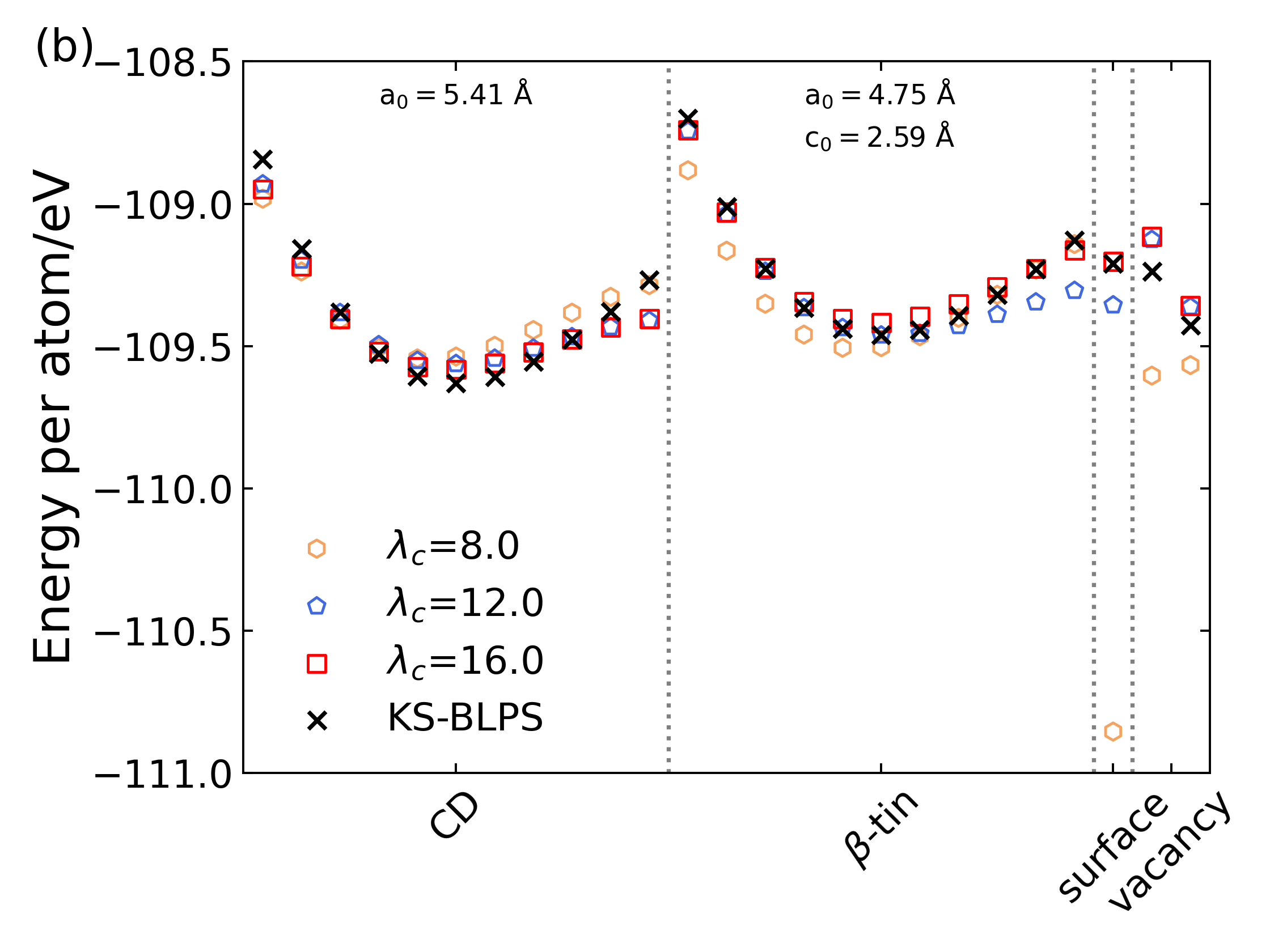}
    % \caption{}
    \label{fig:Fitting_Sia}
    \end{subfigure}

    \caption{Total energies (in eV/atom) of target systems as calculated by using the KS-BLPS and OFDFT methods. In OFDFT calculations, three different TKK kernels with the cutoff being $\lambda_c=8$, 12, and 16 are chosen. 
    The metallic systems are illustrated in (a), including the fcc, hcp, bcc, and sc crystal structures of Al, are compressed and expanded with the equilibrium lattice constant $a_0$ ranging from $0.9a_0$ to $1.1a_0$, and five points are chosen for each structure. 
    In addition, the Al fcc surfaces ((110), (100), and (111) surfaces in turn), as well as the vacancy configurations ($1\times1\times1$, $2\times1\times1$, and $2\times2\times2$ supercells in turn) are chosen.
    For semiconductor systems shown in (b), we choose the cubic diamond (CD) and $\beta$-tin solid phases of Si (compress and expand the unit cell from $0.9a_0$ to $1.1a_0$ to obtain eleven points for each configuration), the CD (100) surface, and the CD vacancy configurations ($1\times1\times1$ and $2\times1\times1$ supercells in turn).
    }
    \label{fig:Fitting}
\end{figure*}

Note that we optimize two types of TKKs, including TKK$^{\mathrm{m}}$ (Fig.~\ref{fig:workflow}(b)) and TKK$^{\mathrm{s}}$ (Fig.~\ref{fig:workflow}(c)) for metals and semiconductors, respectively. The target systems are selected as follows.
For the target metallic systems,
we choose face-centered cubic (fcc), body-centered cubic (bcc), simple cubic (sc), and hexagonal close-packed (hcp) crystal structures of bulk Al.
In addition, the fcc (111), (100), and (110) surfaces of Al are considered.  We also adopt three different supercells ($1\times1\times1$, $2\times1\times1$, and $2\times2\times2$ supercells) of fcc Al, which contain one vacancy, to fit the vacancy formation energies. 
On the other hand, in the target systems for semiconductor systems, we select the cubic diamond (CD) and beta-tin crystal structures of bulk Si. Moreover, we add the (100) surface of the CD Si and two different CD Si supercells containing one vacancy ($1\times1\times1$ and $2\times1\times1$ supercells) in the target systems.

% In particular, the Al fcc, hcp, bcc, and sc crystal structures are compressed and expanded with the equilibrium lattice constant $a_o$ ranging from $0.9a_0$ to $1.1a_0$, and five points are chosen for each structure. 
% In addition, the Al fcc surfaces ((110), (100), and (111) surfaces in turn), as well as the vacancy configurations ($1\times1\times1$, $2\times1\times1$, and $2\times2\times2$ supercells in turn) are chosen as target metallic systems.
% Total energy (in eV/atom) of target semiconductor systems as calculated by KSDFT and OFDFT. From left to right there are CD, $\beta$-tin Si (compress and expand the unit cell from $0.9a_0$ to $1.1a_0$ to obtain eleven points for each configuration), CD (100) surface, and CD vacancy configurations ($1\times1\times1$ and $2\times1\times1$ supercells in turn), calculated by \TKK{s}{}s with several $\lambda_c$. 

%
During the optimization of TKKs, we impose a constrain during the optimization, which ensures the hydrodynamic limit ($\eta=0$) of TKK to be fixed at zero as in the conventional WT KEDF kernel $W(\eta)$.~\cite{16JCTC-Chen}
%
%This is realized by tuning the last coefficient of Spherical Bessel functions at each step.
%
\mc{The optimization is performed for 10 temperatures by using the Metropolis algorithm, and 2,100 and 1,400 steps are carried out for each temperature for \TKK{m}{}s and \TKK{s}{}s KEDFs, respectively.}

%-------------
% Figure 3
%-------------
\begin{figure}[htbp]
    \centering
    
    \begin{subfigure}{0.45\textwidth}
    \centering
    \includegraphics[width=0.95\linewidth]{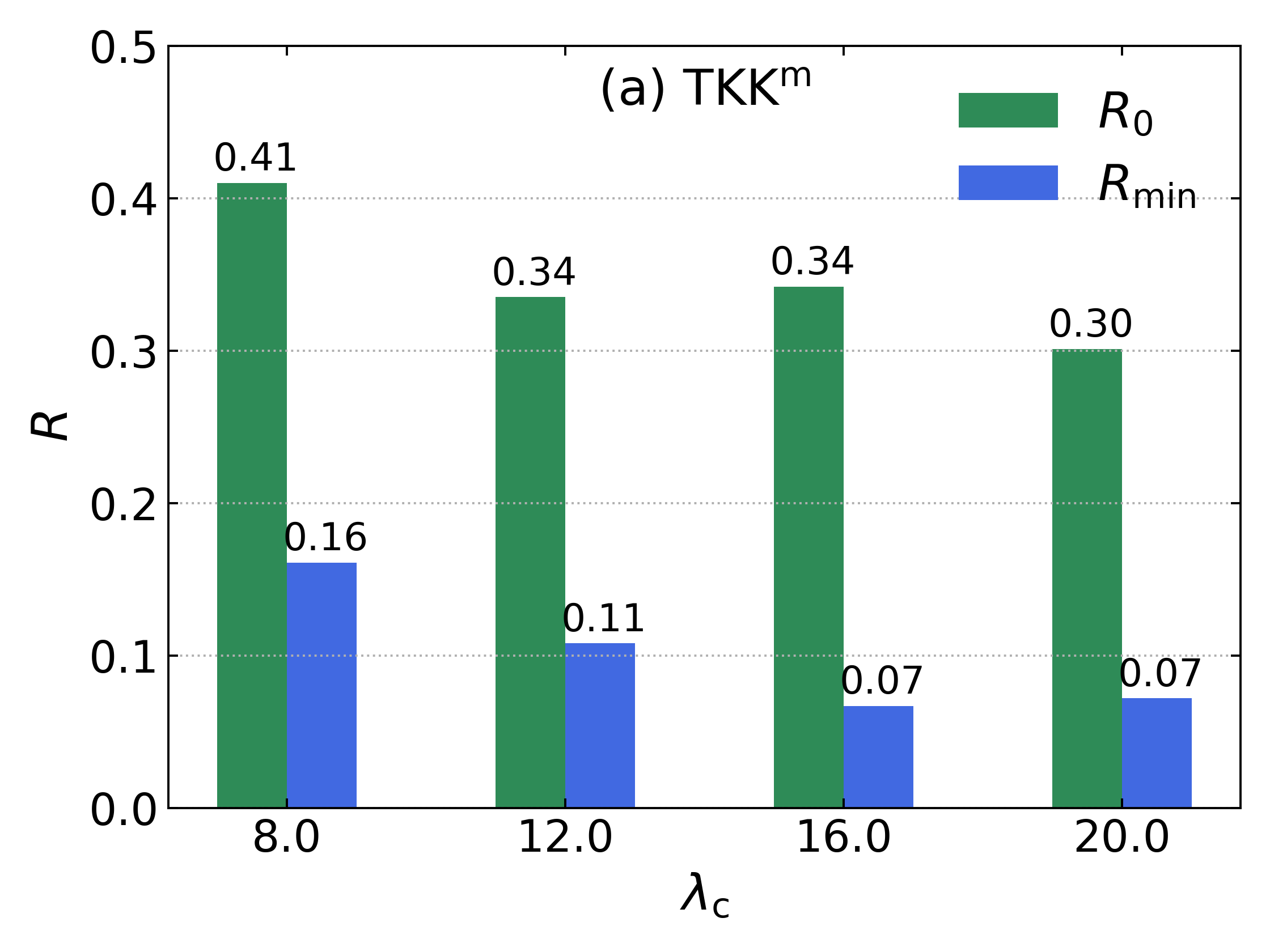}
    \label{fig:Residual_al}
    \end{subfigure}
    \begin{subfigure}{0.45\textwidth}
    \centering
    \includegraphics[width=0.95\linewidth]{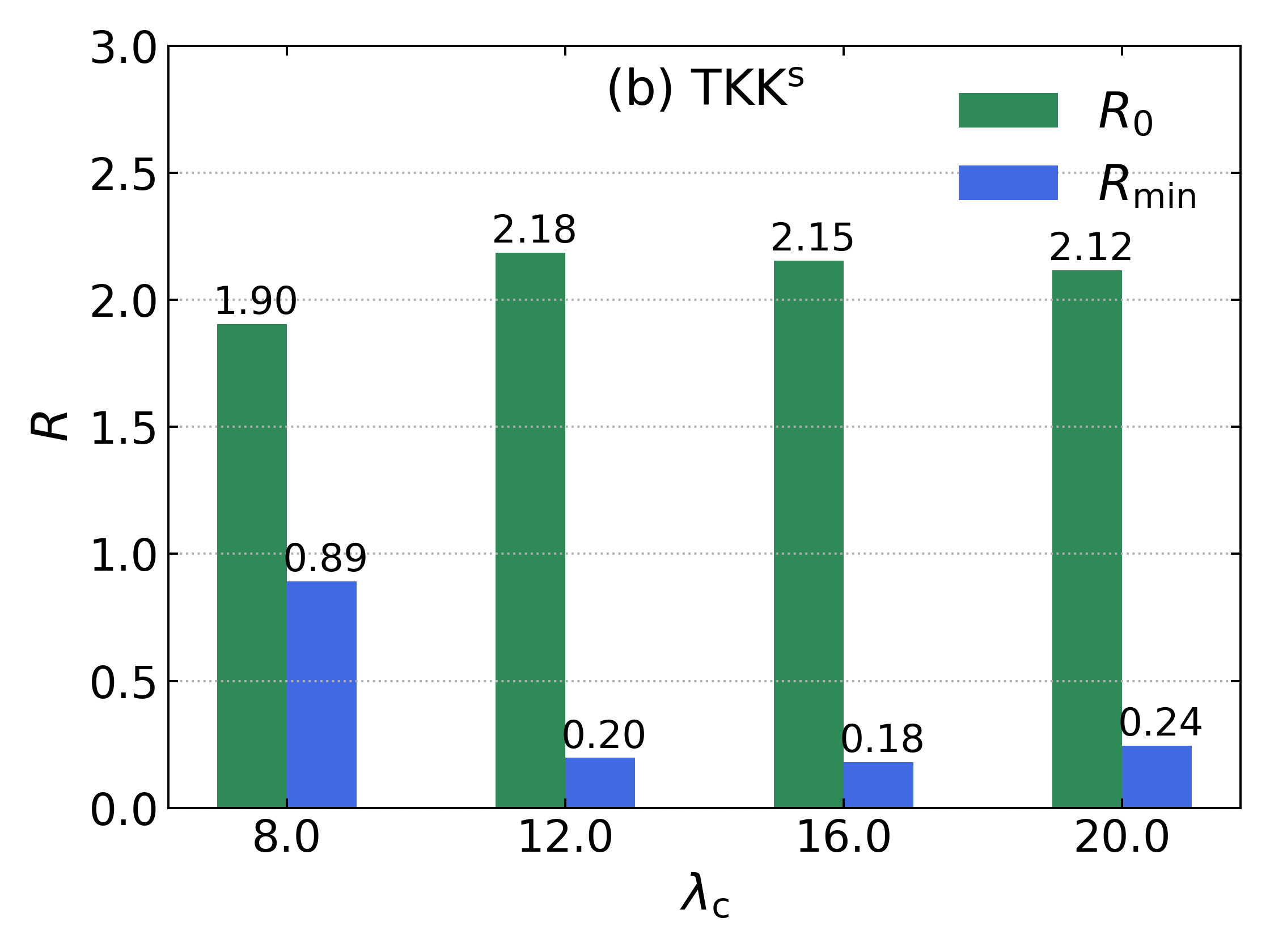}
    \label{fig:Residual_si}
    \end{subfigure}
    
    \caption{Residual values ($R_0$ and $R_\tx{min}$ in Eq.~\ref{eq.residual}) for the target systems before and after the optimization of the truncated KEDF kernel (TKK) with the simulated annealing method. The optimized TKKs are labeled as TKK$^{\rm{m}}$ and TKK$^{\rm{s}}$ for the (a) metallic and (b) semiconductor systems, respectively. The dimensionless radial cutoff of TKKs is $\lambda_c$.
    $R_0$ and ${R_{\tx{min}}}$ are the original residuals before optimization and the minimum residual after optimization, respectively.}
    \label{fig:Residual}
\end{figure}

\section{Numerical Details}

We perform OFDFT and KSDFT calculations by using the PROFESS 3.0~\cite{15CPC-Chen-PROFESS} and ABACUS 2.1.0~\cite{16Li-CMS-ABACUS} packages.
%
%As abovementioned, we select a variety of representative systems for metals and semiconductors to optimize two TKKs. Furthermore, to test the transferability of TKKs for metals, two supercells of Al to compute the stacking fault energy are selected. \recheck{other systems such as semiconductors?} {I have calculated the bulk properties of AlP and GaP with \TKK{s}{} KEDFs, but none of them gave reasonable results.}
%
The plane-wave energy cutoffs utilized in OFDFT and KSDFT calculations for the abovementioned systems, as well as the Monkhorst-Pack $k$-point samplings~\cite{76B-Monkhorst} in KSDFT are listed in Table S1.
In both OFDFT and KSDFT calculations, the local density approximation (LDA)~\cite{81B-Perdew} and the bulk-derived local pseudopotentials (BLPS)~\cite{08PCCP-Huang-BLPS} are used.
In particular, the Gaussian smearing method is used in the KSDFT calculations for metallic systems, with the smearing width being 0.1 eV.
In order to calculate the ground-state bulk properties, we first optimize the crystal structures until the stress tensor elements are below $5\times10^{-7}$\ Hartree/$\text{Bohr}^3$, then compress and expand the unit cell from $0.99a_0$ to $1.01a_0$, where $a_0$ is the equilibrium lattice constant.
Once the energy-volume curve is obtained, bulk modulus $B$ is calculated by fitting Murnaghan’s equation of state.\cite{44-Murnaghan-bulkmodulus}
\mc{In order to assess the error of $B$, we choose two different sets of data for the fcc and hcp Al structures and adopt KSDFT with BLPS and the LDA functional. First, the lattice constant is chosen from 0.990$a_0$ to 1.010$a_0$, The calculated bulk moduli of fcc and hcp Al are 84 and 81 GPa, respectively. Second, we change the range of the lattice constant from 0.995$a_0$ to 1.005$a_0$, and the resulting bulk moduli are 83 and 81 GPa for the fcc Al and hcp Al, respectively. The two sets of data are close, suggesting that our method to estimate the bulk moduli is reasonably accurate.}

We compare the TKK results to those obtained from OFDFT calculations with traditional KEDFs.
In detail, the WT and WGC KEDFs are used for systems involving Al, Li, and Mg. The WT, WGC, and HC KEDFs are adopted for Si systems.
We set $\alpha$=$\frac{5+\sqrt{5}}{6}$, $\beta$=$\frac{5-\sqrt{5}}{6}$ and $\gamma$=2.7 in the WGC KEDF for Al, Li and Mg metals, \mc{which are the optimized parameters of the WGC KEDF as proposed in Ref.~\onlinecite{99B-Wang-nonlocal}}.
In addition, we choose $\alpha$=$\frac{5+\sqrt{5}}{6}$, $\beta$=$\frac{5-\sqrt{5}}{6}$ and $\gamma$=4.2 for the Si systems, \mc{which are optimum for semiconductors, as suggested by Ref.~\onlinecite{08CPC-Ho-profess}}.
The HC KEDF is chosen for Si with the parameters being $\beta=0.65,\ \lambda=0.01$ for the CD structure and $\beta=0.65,\ \lambda=0.0055$ for the $\beta$-tin structure, which is optimum for corresponding systems.~\cite{10B-Huang-nonlocal}
\mc{In all OFDFT calculations, we set the average charge density $\rho_0$ as the average charge density over the whole cell.}

%-----------------------
% surfaces
%-----------------------
The Al fcc (111), (100), and (110) surfaces are respectively tested with 5, 5, and 7 layers of atoms,\SL{~\cite{99B-Wang-nonlocal}} while the Si CD(100) surface is modeled with 9 layers of atoms.~\SL{\cite{10B-Huang-nonlocal}}
In addition, each layer of the above slabs contains one atom, and the lattice vectors are fixed to the equilibrium bulk lattice vectors while the vacuum is set to be larger than 10~${\tx \AA}$.
The surface energy $\sigma$ is defined as
\begin{equation}
    \sigma  = \frac{{{E_{\tx{slab}}} - NE_0}}{{2A}},
\end{equation}
where $E_{\tx{slab}}$ is the total energy of the slab, $E_0$ is the ground energy per atom of bulk fcc Al or CD Si, $N$ is the number of atoms in the slab, and $A$ is the lateral area of the slab.

%--------------------------
% Results 3: point defects 
%--------------------------
The vacancy configurations of fcc Al (CD Si) are set up by removing one atom from a supercell, which is constructed by combining $n$ Al fcc (Si CD) cubic unit cells together in a $n_1 \times n_2 \times n_3$ fashion with $n = n_1 n_2 n_3$.
%{\sout{For brevity, we call such an fcc or a CD supercell as fcc $(n_1 \times n_2 \times n_3)$ or CD $(n_1 \times n_2 \times n_3)$ supercell.}}
%--------------------------
% vacancy formation energy
%--------------------------
Next, the vacancy formation energy $E_{\tx{vf}}$ is calculated via,~\cite{89JP-Gillan-vacancy}
\begin{equation}
    {E_{\tx{vf}}} = E\left( {N - 1,\, 1,\, \frac{N - 1}{N}\Omega } \right) - \frac{N - 1}{N} E(N,\, 0,\, \Omega ),
\end{equation}
where $E\left( {N,\, m,\, \Omega } \right)$ is the total energy for a cell. The parameters $\Omega$, $N$, and $m$ depict the volume, the number of atoms, and the number of defects, respectively.

The Mean Absolute Relative Error (MARE) of property $x$ is defined as 
\begin{equation}
    {\rm{MARE}}=\frac{1}{N}\sum_i^N{|\frac{x_i^{\rm{OF}} - x_i^{\rm{KS}}}{x_i^{\rm{KS}}}|}.
    \label{eq:mare}
\end{equation}
Here $N$ is the number of data points, $x_i^{\rm{OF}}$ and $x_i^{\rm{KS}}$ are obtained from OFDFT and KSDFT, respectively.

%
%------------------------------------
% Results 4: Stacking fault energies
%------------------------------------
The stacking fault energies are calculated with the same setup as in Bernstein and Tadmor’s work\cite{04B-bernstein-stack}. More information, such as the computed stacking fault energies, is shown in Fig.~S1 in SI.

\section{Results and Discussion}

We generate TKKs with different cutoffs, i.e., $\lambda_c=8.0,\,12.0,\,16.0,\,20.0$ for metallic and semiconductor systems; the starting residual function $R_0$ (green bars) and the final residual function $R_\tx{min}$ (blue bars) in terms of different cutoffs are shown in Fig.~\ref{fig:Residual}.
We find that the final residual $R_\tx{min}$ is substantially smaller than the original residual $R_0$, implying that the optimization scheme is effective.
Notably, as $\lambda_c$ increases, the residual decreases first and then increases slightly. This may be caused by the introduction of the long-range part, which enhances the accuracy of TKK. 
However, as the $\lambda_c$ increases, the fitting capability for the two TKKs reaches a saturation point.
%
%Therefore, we expect that if $\lambda_c$ and the number of basis functions are increased simultaneously, better results will be obtained, but based on the consideration of the amount of calculation, we have no further increase the number of basis functions.

%-------------------------------
% Table 1
%-------------------------------
\begin{table}[tbp]
	\centering
	\caption{Correspondence between the real-space distance ($r_1$ and $r_2$) and the real-space cutoff ($\lambda_1$=$2k_\mathrm{F}r_1$ and $\lambda_2$=$2k_\mathrm{F}r_2$) in the fcc Al, hcp Al, CD Si and $\beta$-tin Si crystal systems. Here $r_1$ ($r_2$) depicts the distance between an atom and its nearest neighbor (next-nearest neighbor), and $k_{\tx{F}} = (3\pi^2\rho_0)^{1/3}$ (in $\tx{\AA}^{-1}$) is the Fermi vector with  $\rho_0$ being the average charge density. 
		} 
	\begin{tabularx}{0.99\linewidth}{
			>{\raggedright\arraybackslash}X
			>{\centering\arraybackslash\hsize=1.1\hsize\linewidth=\hsize}X
			>{\centering\arraybackslash\hsize=1.1\hsize\linewidth=\hsize}X
			>{\centering\arraybackslash}X
			>{\centering\arraybackslash}X
			>{\centering\arraybackslash\hsize=0.9\hsize\linewidth=\hsize}X
			>{\centering\arraybackslash\hsize=0.9\hsize\linewidth=\hsize}X}
		\hline\hline
		&$\rho_0\ (\tx{\AA}^{-3})$ &$k_{\tx{F}}\ (\tx{\AA}^{-1})$ &$r_1\ (\tx{\AA})$ &$r_2\ (\tx{\AA})$ &$\lambda_1$ &$\lambda_2$\\
		\hline
		fcc Al  &0.192  &1.785  &2.807  &3.970  &10.020 &14.171\\
		hcp Al  &0.191  &1.782  &2.808  &4.586  &10.006 &16.341\\
		CD Si   &0.202  &1.815  &2.342  &3.824  &8.502  &13.883\\
		$\beta$-tin Si  &0.274  &2.010  &2.462  &2.591  &9.894  &10.412\\
		\hline\hline
	\end{tabularx}
	\label{tab:r2lambda}
\end{table}

\mc{
Furthermore, given that the parameter $\lambda = 2k_{\tx{F}} |{\bf{r}} - {\bf{r'}}|$ is dimensionless, it should be noted that the same real-space cutoff $\lambda_c$ may correspond to varying real-space distances when the Fermi vector $k_{\tx{F}}$ takes different values.}
% the range of the real-space cutoff for TKK depends on the $k_{\tx{F}}$ value in different systems.
To clarify this point, we list the correspondence between the real-space distance and the cutoff $\lambda$ in Table \ref{tab:r2lambda}.
Notably, when it comes to analyzing the behavior of Al and Si systems, it is crucial to take into account the interactions between atoms up to the second nearest neighbor. This is because these interactions can have a significant impact on the overall properties of the systems.
Specifically, TKKs only consider the nearest neighbor atoms for hcp Al when $\lambda_c$ is less than 16 but consider the atoms up to the second neighbors when $\lambda_c$ is larger than 16.
%
% In summary, taking into account interactions up to the second nearest neighbor is essential for accurately analyzing the behavior of Al and Si systems. 
\mc{In summary, \TKK{m}{16} and \TKK{s}{16} take into account interactions up to the second nearest neighbors, whereas other TKKs with $\lambda_c < 16$ exclude these interactions.}

%-------------------------------
% Table 2
%-------------------------------
\begin{table*}[htbp]
	\centering
	\caption{Bulk properties of the fcc, hcp, bcc and sc crystal structures of Al, i.e., the bulk modulus ($B$ in $\tx{GPa}$), the equilibrium volume ($V_0$ in $\tx{\AA}^3$/atom), and the energy of a given system ($E_0$ in eV/atom). 
		The energy ($E_0$) of fcc Al is chosen to be the total energy, while the other energy terms are set as the energy difference between the fcc Al and other structures. 
		The MARE as defined in Eq.~\ref{eq:mare} is obtained by comparing OFDFT to KS-BLPS results. 
Both KSDFT and OFDFT calculations are performed with the use of bulk-derived pseudopotentials (BLPS). 
For OFDFT calculations, we use the WT KEDF and the TKKs for metals with different cutoffs $\lambda_c$ (labeled as TKK$^{m}_{\lambda_c}$).
Some of the KS-BLPS and WGC KEDF data are taken from Ref.~\onlinecite{08PCCP-Huang-BLPS}.
The experimental data of bulk moduli and equilibrium volumes for fcc Al are shown for comparison. 
        }
	\begin{tabularx}{0.9\linewidth}{
			>{\raggedright\arraybackslash}X
			>{\raggedright\arraybackslash}X
			>{\centering\arraybackslash}X
			>{\centering\arraybackslash}X
			>{\centering\arraybackslash}X
			>{\centering\arraybackslash}X
			>{\centering\arraybackslash}X}
		\hline\hline
		&       &fcc    &hcp    &bcc    &sc &MARE\\
		\hline
		$B\ (\tx{GPa})$	&KS-BLPS (this work)	&\mc{84}	&\mc{81}	&\mc{77}	&\mc{66}   &-\\
            &KS-BLPS~\cite{08PCCP-Huang-BLPS}    &\mc{84}   &\mc{81}   &\mc{76}   &\mc{64}   &-\\
		% &WGC	        &81.0	&79.4	&74.9	&62.2   &3.20\\
		&WGC\cite{08PCCP-Huang-BLPS}    &\mc{81}   &\mc{80}   &\mc{75}   &\mc{62}   &3.11$\%$\\
		&WT	            &\mc{85}	&\mc{83}	&\mc{77}	&\mc{65}   &1.72$\%$\\
		%&truncated WT	&84.3	&83.9	&79.7	&60.5   &4.13\\
		%&\TKK{m}{20}	&86.5	&84.5	&76.3	&66.1   &2.22\\
		&\TKK{m}{16}	&\mc{87}	&\mc{85}	&\mc{76}	&\mc{69}   &3.99$\%$\\
		&\TKK{m}{12}	&\mc{89}	&\mc{88}	&\mc{81}	&\mc{68}   &5.88$\%$\\
            &\TKK{m}{8}	    &\mc{80}	&\mc{80}	&\mc{75}	&\mc{47}   &9.01$\%$\\
		&Exp.\cite{79JPCS-Tallon-al_modulus_exp}     &76.2\\
		\\
		$V_0\ (\tx{\AA}^3)$ &KS-BLPS (this work)  &15.644 &15.741 &16.084 &18.797  &-\\
            &KS-BLPS~\cite{08PCCP-Huang-BLPS}    &15.623   &15.767  &16.063   &18.825   &-\\
		% &WGC    &15.632	&15.697	&15.887	&19.226 &0.97\\
		&WGC\cite{08PCCP-Huang-BLPS}    &15.632 &15.764 &15.887 &19.223 &0.93$\%$\\
		&WT	            &15.821	&15.928	&16.223	&18.774 &0.83$\%$\\
		%&truncated WT   &16.054	&16.056	&16.263	&19.213 &1.99\\
		%&\TKK{m}{20}	&$15.637$	&15.735	&16.114	&18.612 &0.31\\
		&\TKK{m}{16}	&15.646	&15.712	&16.117	&18.568 &0.41$\%$\\
		&\TKK{m}{12}	&15.729	&15.777	&16.107	&18.507 &0.61$\%$\\
        &\TKK{m}{8}    &16.005	&15.997	&16.192	&20.130 &2.92$\%$\\
		&Exp.\cite{71ACSA-Straumanis-al_lattice_exp}  &16.363\\
		\\
		$E_0\ (\tx{eV})$	&KS-BLPS (this work)	&$-57.949$	&0.027	&0.087	&0.361  &-\\
            &KS-BLPS~\cite{08PCCP-Huang-BLPS}    &$-57.955$   &0.038   &0.087   &0.362   &-\\
		&WGC\cite{08PCCP-Huang-BLPS}    &$-57.941$  &0.018  &0.079  &0.354  &0.00$\%$\\
		% &WGC	        &$-57.940$ 	&0.017	&0.078	&0.352  &0.00\\
		&WT	            &$-57.934$ 	&0.020	&0.078	&0.335  &0.02$\%$\\
		%&truncated WT	&$-57.913$	&0.002	&0.046	&0.359  &0.04\\
		%&\TKK{m}{20}	&$-57.930$	&0.014	&0.082	&0.339  &0.02\\
		&\TKK{m}{16}	&$-57.949$	&0.021	&0.080	&0.366  &0.01$\%$\\
		&\TKK{m}{12}	&$-57.900$	&0.000	&0.043	&0.280  &0.05$\%$\\
		&\TKK{m}{8}	&$-57.914$	&0.000	&0.057	&0.293  &0.04$\%$\\
		\hline\hline
	\end{tabularx}
	\label{tab:Al_bulk}
\end{table*}

%-------------------------------
% Table S3: Li and Mg
%-------------------------------
% \renewcommand{\arraystretch}{1}
\begin{table*}[htbp]
	\centering
	\caption{\mc{OFDFT and KSDFT results for bulk modulus ($B$ in $\tx{GPa}$), equilibrium volume ($V_0$ in $\tx{\AA}^3$ per atom), and total energy ($E_0$ in eV per atom) of various solid phases of Li and Mg. The last column is MARE (\%). The equilibrium total energies of bcc Li and hcp Mg are given, while the energy differences are shown for other structures. All results of KSDFT and WGC KEDF for Mg are taken from Ref.\onlinecite{08PCCP-Huang-BLPS}.}}
	\begin{tabularx}{0.9\linewidth}{
			>{\raggedright\arraybackslash}X
			>{\raggedright\arraybackslash}X
			>{\centering\arraybackslash}X
			>{\centering\arraybackslash}X
			>{\centering\arraybackslash}X
			>{\centering\arraybackslash}X
			>{\centering\arraybackslash}X}
		\hline\hline
		Li & &bcc	&fcc	&sc	    &CD    &MARE\\
		\hline
		$B\ (\tx{GPa})$   &KS-BLPS	&\mc{17}	&\mc{17}	&\mc{17}	&\mc{12}   &-\\
		%&truncated WT	&17.0	&17.4	&17.4	&12.3   &0.63\\
		&WGC	&\mc{17}	&\mc{17}	&\mc{17}	&\mc{12}   &0.15\\
		&WT	    &\mc{17}	&\mc{17}	&\mc{17}	&\mc{12}   &0.15\\
            &\TKK{m}{16}	&\mc{17}	&\mc{18}	&\mc{17}	&\mc{12}   &0.70\\
		% \\
		$V_0\ (\tx{\AA}^3)$ &KS-BLPS	&18.767	&18.693	&19.441	&21.929 &-\\
		%&truncated WT	&18.746	&18.676	&19.555	&21.995 &0.27\\
		&WGC	&18.810	&18.728	&19.528	&21.956 &0.25\\
		&WT	    &18.796	&18.714	&19.495	&21.980 &0.19\\
            &\TKK{m}{16}	&18.699	&18.628	&19.474	&21.989 &0.29\\
		% \\
		$E_0\ (\tx{eV})$	&KS-BLPS	&$-7.599$	&$-0.0004$	&0.139	&0.538  &-\\
		%&truncated WT	&$-7.598$	&$-0.001$	&0.144	&0.538  &0.03\\
		&WGC	&$-7.595$	&$-0.002$	&0.140	&0.535  &0.04\\
		&WT	    &$-7.595$	&$-0.002$	&0.140	&0.536  &0.04\\
            &\TKK{m}{16}	&$-7.599$	&$-0.003$	&0.145	&0.543  &0.05\\
		\hline
		Mg		            &       &hcp	&fcc	&bcc	&sc &MARE\\
		\hline
		$B\ (\tx{GPa})$         &KS-BLPS\cite{08PCCP-Huang-BLPS}  &\mc{38}   &\mc{38}   &\mc{37}   &\mc{29}   &-\\
		% &KS-BLPS	&38.2	&37.6	&37.1	&29.2   &-\\
		&WGC\cite{08PCCP-Huang-BLPS}    &\mc{36}   &\mc{36}   &\mc{36}   &\mc{28}   &4.15\\ %4.30\\
		% 	&WGC	&36.6	&36.2	&35.7	&27.9   &3.88\\ %4.03\\
		&WT	    &\mc{37}	&\mc{36}	&\mc{36}	&\mc{29}   &2.54\\ %2.70\\
		%&truncated WT	&36.6	&36.5	&35.8	&30.3   &3.61\\ %3.60\\
		% 	&\TKK{m}{20}	&37.2	&37.0	&35.7	&28.2   &2.70\\ %2.85\\
            &\TKK{m}{16}	&\mc{37}	&\mc{37}	&\mc{36}	&\mc{29}   &2.02\\ %2.18\\
		% 	&\TKK{m}{12}	&36.9	&37.2	&35.5	&30.3   &3.16\\ %3.14\\
		% 	&\TKK{m}{8} 	&36.4	&36.2	&35.7	&28.8   &3.24\\ %3.39\\
		% \\
		$V_0\ (\tx{\AA}^3)$ &KS-BLPS\cite{08PCCP-Huang-BLPS}  &21.176 &21.363 &21.393 &24.929 &-\\
		% &KS-BLPS	&21.165	&21.358	&21.349	&24.908 &-\\
		&WGC\cite{08PCCP-Huang-BLPS}    &21.616 &21.465 &21.534 &25.036 &0.91\\ %1.00\\
		% &WGC	&21.397	&21.481	&21.551	&25.052 &0.71\\ %0.80\\
		&WT	    &21.358	&21.533	&21.590	&25.006 &0.72\\ %0.81\\
		%&truncated WT	&21.833	&21.822	&22.015	&24.826 &2.14\\ %2.19\\
		% 	&\TKK{m}{20}	&21.187	&21.252	&21.360	&25.064 &0.32\\ %0.32\\
            &\TKK{m}{16}	&21.246	&21.312	&21.384	&24.933 &0.16\\ %0.22\\
		% 	&\TKK{m}{12}	&21.354	&21.314	&21.594	&24.672 &0.76\\ %0.80\\
		% 	&\TKK{m}{8}	&21.802	&21.795	&21.948	&25.019 &1.98\\ %2.08\\
		% \\
		$E_0\ (\tx{eV})$	&KS-BLPS\cite{08PCCP-Huang-BLPS}  &$-24.678$  &0.011  &0.033  &0.370  &-\\
		% &KS-BLPS	&$-24.671$  &0.011	&0.031	&0.370  &-\\
		&WGC\cite{08PCCP-Huang-BLPS}    &$-24.651$  &0.006  &0.024  &0.351  &0.08\\ %0.05\\
		% 	&WGC	&$-24.647$	&0.006	&0.024	&0.351  &0.09\\ %0.07\\
		&WT	    &$-24.654$	&0.010	&0.032	&0.352  &0.08\\ %0.05\\
		%&truncated WT	&$-24.629$	&0.001	&0.021	&0.320  &0.13\\ %0.12\\
		% 	&\TKK{m}{20}	&$-24.657$	&$0.000$	&0.034	&0.358  &0.06\\ %0.04\\
            &\TKK{m}{16}	&$-24.652$	&$0.007$	&0.027	&0.337  &0.08\\ %0.06\\
		% 	&\TKK{m}{12}	&$-24.654$	&$-0.006$	&0.039	&0.333  &0.08\\ %0.06\\
		% 	&\TKK{m}{8}	&$-24.630$	&$0.002$	&0.021	&0.326  &0.13\\ %0.11\\
		% \hline
%		Mg Al alloys    &   &   &$\beta''\tx{-}\tx{Al}_3\tx{Mg}$    &   &$\tx{L1}_2\ \tx{Mg}_3\tx{Al}$   &\\
%		\hline
%		$B\ (\tx{GPa})$         &\TKK{m}{16}   &&64.0  &&38.9  &\\
%		&truncated WT   &&62.2  &&39.3  &\\
%		&WT             &&62.8  &&39.3  &\\
%		&WGC            &&66.8  &&47.3  &\\
%		&KS-BLPS          &&67.2  &&47.5  &\\
%		\\
%		$V_0\ (\tx{\AA}^3)$ &\TKK{m}{16}   &&16.956  &&20.138  &\\
%		&truncated WT   &&17.450  &&20.645  &\\
%		&WT             &&17.152  &&20.270  &\\
%		&WGC            &&16.786  &&19.469  &\\
%		&KS-BLPS          &&16.773  &&19.547  &\\
%		\\
%		${\Delta}E_\tx{f}\ (\tx{meV})$  &\TKK{m}{16}   &&11.4      &&$-25.0$   &\\
%		&truncated WT   &&27.4      &&$-27.1$   &\\
%		&WT             &&32.8      &&$-19.1$   &\\
%		&WGC            &&$-6.2$    &&5.8       &\\
%		&KS-BLPS          &&$-9.0$    &&$-15.7$   &\\
		\hline\hline
	\end{tabularx}
	\label{tab:LiMg_bulk}
\end{table*}

\subsection{\mc{Simple Metals}}

Table \ref{tab:Al_bulk} lists the bulk properties of the fcc, hcp, bcc, and sc crystal structures of Al as obtained from KSDFT and various kinetic energy functionals adopted in OFDFT. 
The bulk properties include the bulk modulus, the equilibrium volume, and the total energy.
When compared to the experimental data, we find that the KSDFT method with the usage of the BLPS yields a slightly larger bulk modulus and a smaller equilibrium volume for fcc Al, but the results are reasonable. 
In addition, both KSDFT and OFDFT (the WGC and WT KEDFs) calculations yield similar bulk properties for the four phases of Al, including the prediction of the fcc structure to be the most stable solid phase among the four solid structures.

It is worth mentioning that the energy difference between the hcp and fcc structures is as small as 0.020, 0.018, and 0.027 eV/atom, which is obtained from the WT, WGC, and KS-BLPS calculations, respectively.
Notably, the TKK$^{\rm{m}}_{\lambda_c}$ in OFDFT exhibits different levels of accuracy for the bulk properties of solid Al phases in terms of the dimensionless radius cutoff $\lambda_c$.
In general, a higher accuracy of TKK$^{\rm{m}}_{\lambda_c}$ is obtained while $\lambda_c$ increases from 8 to 16, approaching the accuracy of the WT/WGC KEDF.
Interestingly, we notice that the energy difference between the hcp and fcc structures as obtained from \TKK{m}{8}, \TKK{m}{12}, and \TKK{m}{16} is 0.000, 0.000, and 0.021 eV/atom, respectively.
The results indicate that the TKK$^{\tx{m}}_{\lambda_c}$ with $\lambda_c=8$ or $12$ predicts the same energy for fcc and hcp structures.
\mc{
In addition, we note that the small energy difference between the fcc and hcp Al structures predicted by KSDFT-BLPS with the LDA functional and KSDFT-BLPS with the PBE functional are quite close, which are 0.027 and 0.025 eV/atom, respectively.}

As explained in Table~\ref{tab:r2lambda}, 
the results suggest that the relatively short-ranged TKKs of $\lambda_c=8$ or $12$, which involve only the nearest neighbors, are not able to distinguish the subtle energy difference between the fcc and hcp structures of Al. The reason is the two structures have similar local structures, which are closely packed planes of atoms, and they own the same atomic packing factor of 0.74 and the same coordination number of 12.
Notably, we emphasize that the \TKK{m}{16} KEDF with $\lambda_c=16$, which involves the second nearest neighbors, yields a satisfactory value of 0.021 eV/atom for the energy difference between the hcp and fcc structures. In addition, the bulk moduli and equilibrium volumes of the four structures of Al, as obtained from the \TKK{m}{16} KEDF, match better with the KS-BLPS data as compared to those obtained from \TKK{m}{8} and \TKK{m}{12}. This can be seen by comparing the MARE.

\begin{figure}[tbp]
    \centering
    
    \begin{subfigure}{0.48\textwidth}
    \centering
    \includegraphics[width=0.98\linewidth]{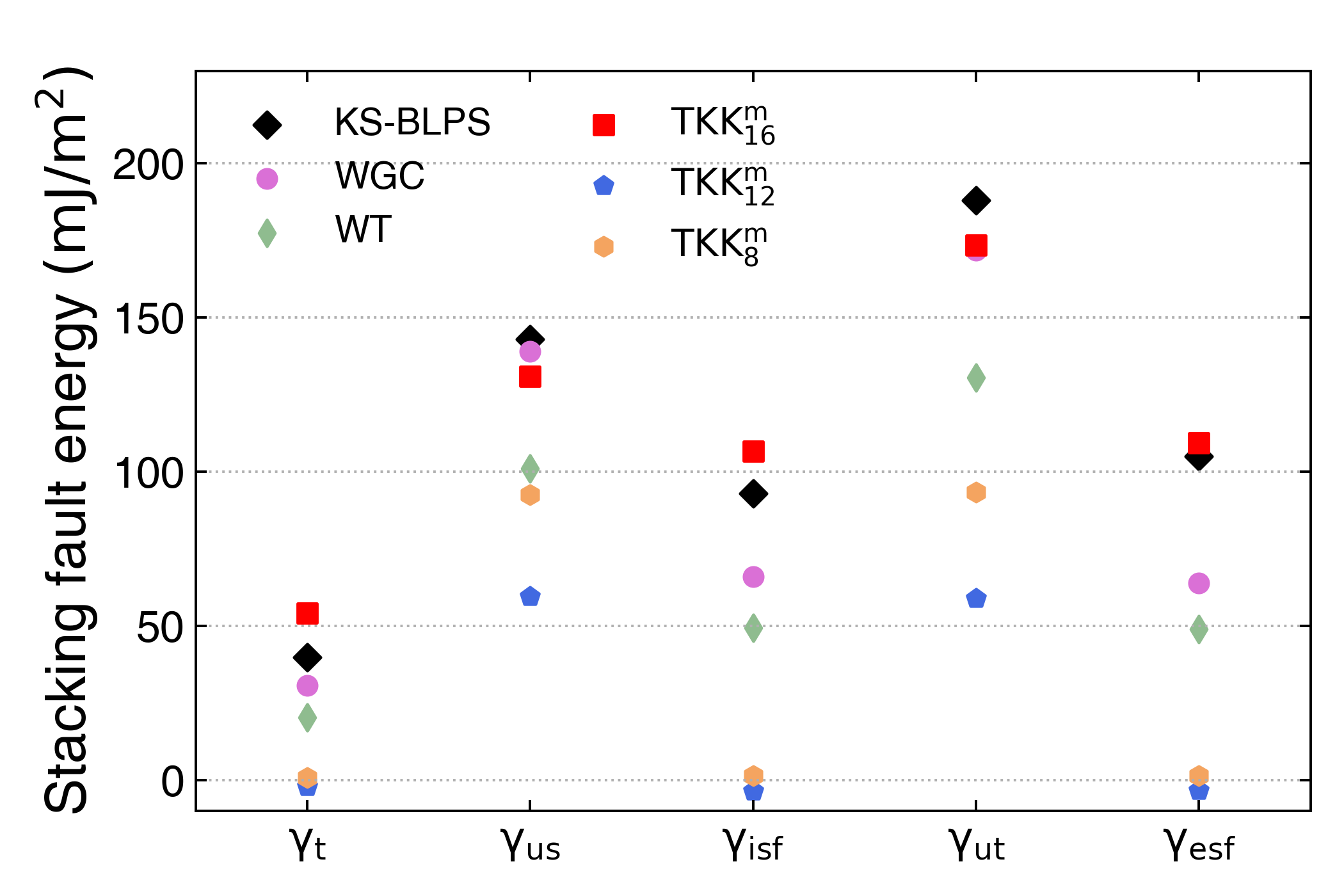}
    \label{fig:Stack_Ala}
    \end{subfigure}
    
    \caption{Stacking fault energies (in $\rm{mJ/m^2}$) of Al as obtained from KS-BLPS~\cite{08PCCP-Huang-BLPS} and OFDFT calculations. The stacking fault energies include the twinning energy $\gamma_\tx{t}$, the unstable stacking fault energy $\gamma_\tx{us}$, the intrinsic stacking fault energy $\gamma_\tx{isf}$, the unstable twinning energy $\gamma_\tx{ut}$, and the extrinsic stacking fault energy $\gamma_\tx{esf}$ of fcc Al. The definition and explanation of the above five stacking fault energies can be found in Ref.~\onlinecite{04B-bernstein-stack}.}
    \label{fig:Stack_Al}
\end{figure}

To validate the transferability of TKKs, we perform OFDFT calculations of the stacking fault energies of fcc Al, which are crucial mechanical properties of metals, and the results are shown in Fig.~\ref{fig:Stack_Al}.
%To further test the effect of long-range part on TKK's discernibility,  because the atoms near the stacking fault plane are arranged in hcp configuration.
Notably, most OFDFT calculations yield smaller stacking fault energies than KSDFT.
In particular, the TKK KEDFs with a small $\lambda_c$ (8 and 12) yield incorrect intrinsic stacking fault energy $\gamma_\tx{isf}$ and extrinsic stacking fault energy $\gamma_\tx{esf}$ close to zero. This can be explained by their inability to distinguish between the hcp and fcc crystal structures since the atoms near the stacking fault plane of fcc Al are arranged in the hcp configuration.
Interestingly, when the cutoff of TKK increases to the second nearest neighbor, we observe that the \TKK{m}{16} KEDF yields reasonable stacking fault energies, which are even better than those obtained from the WT and WGC KEDFs.
%
%As $\lambda_c$ further increases, base functions become flat and stacking fault energies decrease gradually.
%As a result, when base functions are suitable, 
Since the \TKK{m}{16} function has a larger real-space cutoff than the \TKK{m}{8} and \TKK{m}{12} functions, we conclude that the long-range part in the real-space form of the TKK function is important 
to distinguish the energy difference between the fcc and hcp structures of Al, which is crucial to obtain reasonable stacking fault energies.

%-------------
% Figure 4
%-------------

%-------------------------------
% Table 3
%-------------------------------
\begin{table*}[ht]
	\begin{threeparttable}
		\centering
		\caption{Surface energies ($\sigma$, in $\rm{mJ/m^2}$) and vacancy formation energies ($E_{\rm{vf}}$, in eV) of fcc Al and CD Si. The BLPS of Al and Si are used in both KSDFT and OFDFT calculations.}
		\begin{tabularx}{0.9\linewidth}{
				>{\centering\arraybackslash}X
				>{\centering\arraybackslash}X
				>{\centering\arraybackslash}X
				% >{\centering\arraybackslash}X
				% >{\centering\arraybackslash}X
				>{\centering\arraybackslash}X
				>{\centering\arraybackslash}X
				>{\centering\arraybackslash}X
				>{\centering\arraybackslash}X
				>{\centering\arraybackslash}X}
			\hline\hline
			Al          &Systems    &KS-BLPS    &WGC    &WT &\TKK{m}{16} &\TKK{m}{12}   &\TKK{m}{8}\\
			\hline
			%     % KS-BLPS(this work)	&1011	&1110	&1225\\
			$\sigma$    &Al fcc (111)   &1010\tnote{a}   &1176    &1808     &919    &81  &584\\
			&Al fcc (100)   &1104\tnote{a}   &1373   &1971    &1117    &394     &848\\
			&Al fcc (110)   &1212\tnote{a}   &1378   &1996    &1276    &485   &813\\
			\\
			$E_{\rm{vf}}$\tnote{b}  &Al $1\times1\times1$  &0.796  &0.706  &1.237  &0.931  &0.725  &0.574\\
			&Al $2\times1\times1$  &0.757  &0.740  &1.347  &0.905  &0.786  &0.751\\
			&Al $2\times2\times1$  &0.747  &0.809  &1.407  &0.769  &0.794  &0.799\\
			&Al $2\times2\times2$  &0.794  &0.874  &1.447  &0.592  &0.763  &0.791\\
			\hline
			Si          &Systems    &KS-BLPS &HC &WT &\TKK{s}{16} &\TKK{s}{12} &\TKK{s}{8}\\
			\hline
			$\sigma$    &Si CD (100)    &2062  &$2548$  &$-7824$  &$2228$  &$1307$  &$-6172$\\
            $E_{\rm{vf}}$\tnote{c}  &Si $1\times1\times1$ &$2.735$ &$2.651$ &$-0.572$ &$3.277$    &$3.081$ &$-0.552$\\
            &Si $2\times1\times1$ &$3.026$ &$2.313$ &$-0.453$ &$3.367$ &$3.024$ &$-0.551$\\
            &Si $2\times2\times2$ &$3.240$ &$1.445$ &$-0.346$ &$3.583$ &$3.143$ &$-0.575$\\
			\hline\hline
		\end{tabularx}
		\label{tab:Surface_vacancy}
		\begin{tablenotes}
			\item [a] Ref.~\onlinecite{08PCCP-Huang-BLPS}.
			\item [b] The experimental value is 0.66 eV.\cite{75B-Triftshauser-al_vacancy_exp}
			\item [c] The experimental value is 3.6 eV.\cite{64PR-Watkins-si_vacancy_exp1, 86L-Dannefaer-si_vacancy_exp2}
		\end{tablenotes}
	\end{threeparttable}
\end{table*}

% Al surface energies
As listed in Table~\ref{tab:Surface_vacancy}, the surface energies of the Al fcc (100), (110), and (111) surfaces are computed by both OFDFT and KSDFT with the usage of BLPS. 
We find that the WT KEDF significantly overestimates the surface energies as compared to the KSDFT data.
In detail, the KSDFT predicts the surface energies of Al to be 1010, 1104, and 1212 $\rm{mJ/m^2}$ for the  fcc (111), (100), and (110) surfaces, respectively;
the WT KEDF yields surface energy of 1808, 1971, and 1996 $\rm{mJ/m^2}$ for the fcc (111), (100), and (110) surfaces, respectively.
Furthermore, the WGC KEDF largely improves the data, giving rise to a surface energy of 1176, 1373, and 1378 $\rm{mJ/m^2}$ for the fcc (111), (100), and (110) surfaces, respectively.
In terms of the TKKs with different cutoffs, the \TKK{m}{8} and \TKK{m}{12} KEDFs predict significantly smaller values for the surface energies of fcc Al as compared to the KSDFT data, which may be due to the reason of the short-ranged features of the two kinetic energy kernels in describing the kinetic energies of electrons.
Notably, the \TKK{m}{16} KEDF not only predicts the surface energies of fcc Al close to the KSDFT results but also yields the correct energy orderings for the three surfaces of fcc Al.

% %-------------
% % Figure 5
% %-------------
% \begin{figure}[tbp]
%     \centering
    
%     \begin{subfigure}{0.48\textwidth}
%     \centering
%     \includegraphics[width=0.98\linewidth]{Figs/Fig5-a.png}
%     % \caption{\label{fig:Vacancy_al}}
%     \label{fig:Vacancy_al}
%     \end{subfigure}
%     \begin{subfigure}{0.48\textwidth}
%     \centering
%     \includegraphics[width=0.98\linewidth]{Figs/Fig5-b.png}
%     % \caption{}
%     \label{fig:Vacancy_si}
%     \end{subfigure}
    
%     \caption{Vacancy energies (in eV) of (a) fcc Al and (b) CD Si as computed from KS-BLPS and OFDFT methods with various kinds of KEDFs. The system sizes are {3, 7, 15, 31, 107, 255, 499, 863, 1371 atoms for fcc Al, and 7, 15, 63, 215, 511, 999, 1727, 2743 atoms for CD Si}.}
%     \label{fig:Vacancy}
% \end{figure}

%-------------
% Figure 5
%-------------
\begin{figure*}[htbp]
	\centering
	\includegraphics[width=1.0\textwidth]{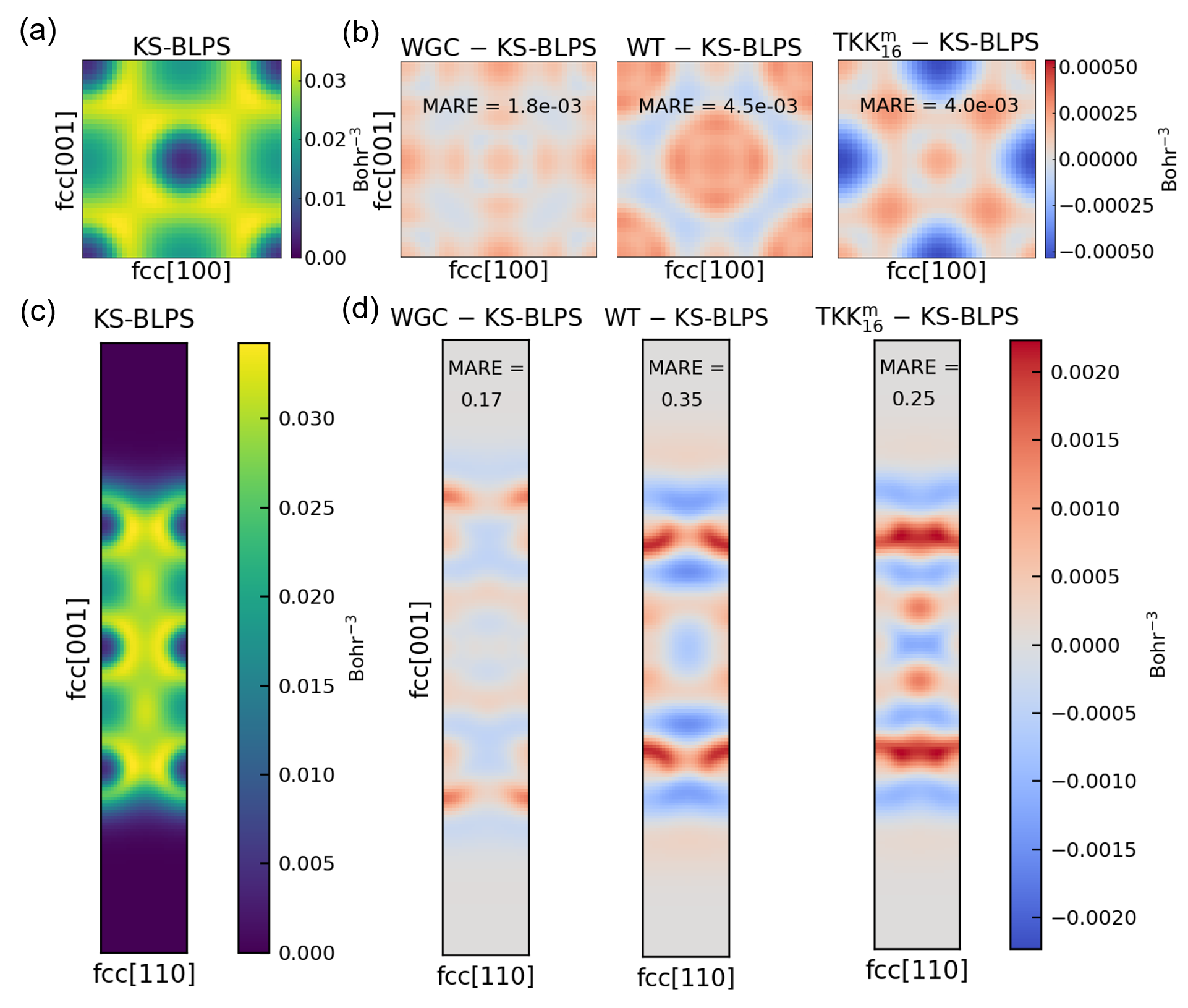}\\
	\caption{
	(a) Electron density profile of the (010) crystal surface inside bulk fcc Al, which is obtained from KS-BLPS calculations.
    (b) Electron density differences of the fcc Al (010) crystal surface between OFDFT and KSDFT calculations. 
    (c) Electron density profile on the longitudinal section of the Al fcc (100) surface, i.e., the fcc $(1\bar{1}0)$ surface. Results are obtained from KSDFT calculations. Fig.~S3(a) of SI shows the slab configuration.
    (d) Electron density differences of the Al fcc (100) surface between OFDFT and KSDFT calculations. 
    We perform KSDFT calculations to obtain the equilibrium configuration used in the above calculations. 
    The HC, WT, and \TKK{s}{16} KEDFs are adopted in OFDFT.
    The MAREs, as defined in Eq.~\ref{eq:mare}, of density differences shown in the figures are calculated from the whole electron density in the \mc{cell for bulk fcc Al and Al fcc (100) surface}.
	}\label{fig:fccAl_den}
\end{figure*}

%-------------
% Figure 6
%-------------
\begin{figure*}[htbp]
	\centering
	\includegraphics[width=1.0\textwidth]{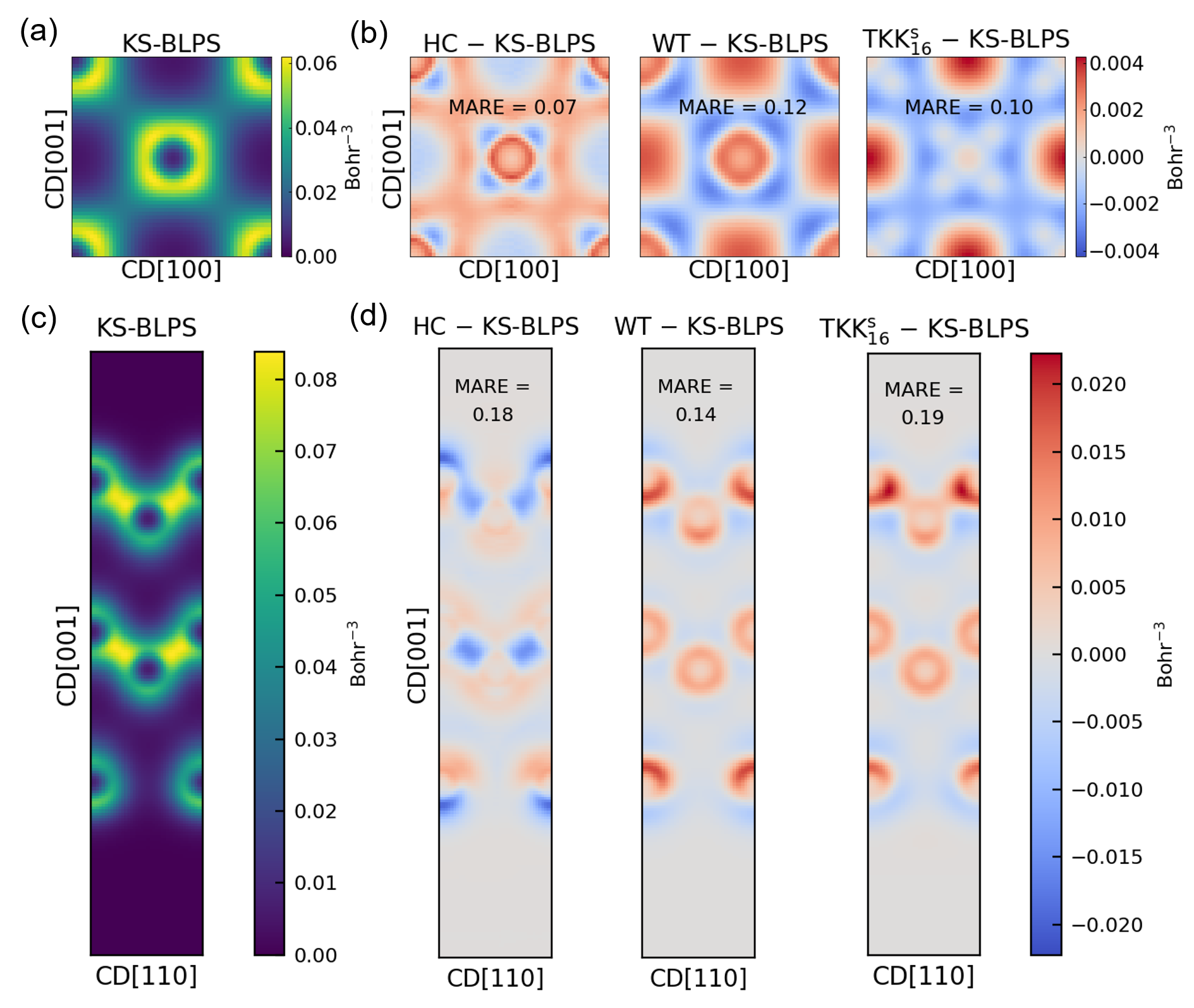}\\
	\caption{
	(a) Electron density profile of the (010) crystal surface inside bulk CD Si as obtained from KSDFT.
    (b) Electron density differences of the CD Si (010) crystal surface between OFDFT and KSDFT calculations.
    (c) Electron density on the longitudinal section of the Si CD (100) surface, i.e., the CD $(1\bar{1}0)$ surface, as obtained from KSDFT calculation. The slab configuration is shown in Fig.~S3(b) of SI.
    (d) Electron density differences of the Si CD (100) surface between OFDFT and KSDFT calculations. 
    All of the calculations are performed in the equilibrium configuration obtained by KSDFT. 
    The HC, WT, and \TKK{s}{16} KEDFs are used in OFDFT.
    The MAREs, as defined in Eq.~\ref{eq:mare}, of density differences are shown in the figures. The MAREs are calculated based on the whole electron density in the \mc{cell for bulk CD Si and Si CD (100) surface}.
	}\label{fig:cdSi_den}
\end{figure*}

Table~\ref{tab:Surface_vacancy} lists the vacancy formation energies of Al. 
First of all, the results obtained by the WT KEDF are substantially larger than the KSDFT data.
For example, the vacancy formation energy of a 2$\times$2$\times$2 cell is 1.447 and 0.794 \mc{eV} from the WT and KS-BLPS calculations, respectively. The WGC KEDF is able to improve the value to 0.874 eV, which is closer to KSDFT.
Second, we observe that all of the three \TKK{m}{} KEDFs yield a higher level of accuracy for the vacancy formation energies than the WT KEDF. Among them, the \TKK{m}{12} KEDF performs best for all of the system sizes studied.  Fig.~S2(a) in SI shows the convergence behaviors of the vacancy formation energy with respect to the system size. We find the \TKK{m}{12} and \TKK{m}{8} KEDFs exhibit \mc{a higher accuracy} in predicting the vacancy formation energy in large systems compared to \TKK{m}{16} KEDF, which may be attributed to the lack of sufficiently large supercells with vacancies in the target systems.

Figs.~\ref{fig:fccAl_den}(a) and (c) respectively illustrate the electron densities of bulk fcc Al and its (100) surface as obtained from KSDFT and OFDFT calculations, and we choose the (010) surface of bulk Al to plot the electron density profile.
For the bulk Al, the electron density differences between OFDFT and KSDFT calculations and the associated MAREs are displayed in Fig.~\ref{fig:fccAl_den}(b). We observe that the WGC, WT, and \TKK{m}{16} KEDFs are capable of reproducing the ground state charge density obtained by KSDFT, with MAREs on the same order of $0.001$ (0.0018, 0.0045, 0.0040 in turn).
On the other hand, when dealing with the Al fcc (100) surface, \mc{we found the small electron density in vacuum may result in a large contribution to the MARE as defined in Eq.~\ref{eq:mare}. Therefore, we did not calculate the contributions to MARE for a selected length of vacuum (5.2 \AA) in the Al slab. The resulting MAREs of Al fcc (100) surface are 0.17, 0.35, and 0.25 for the WGC, WT, and \TKK{m}{16} KEDFs, respectively. We observe that the MAREs of fcc (100) surface are still two orders of magnitude larger than those of Al fcc bulk system.
This phenomenon can be rationalized by the fact that the KEDFs are based on Lindhard response function, which is suitable for describing Al bulk systems with electron density distribution similar to the uniform electron gas. However, when applied to Al surfaces, the electron density changes rapidly around the surfaces and substantially deviates from the uniform electron gas, resulting in a large MARE.}
%
%This is expected because the KEDFs are based on the uniform electron gas model, so that they cannot handle the rapid changes in charge density that occur near surfaces.
Still, we find the \TKK{m}{16} KEDF achieves a similar accuracy when compared to the WGC and WT KEDFs, demonstrating that the \TKK{m}{16} KEDF is able to accurately describe the electronic structure of metallic systems.

% transferability
\mc{
In order to assess the transferability of \TKK{m}{16} KEDF, we conduct bulk property calculations for various solid phases of Li and Mg.
The results are compared with those obtained from WGC, WT KEDFs, and KSDFT, which are presented in Table~\ref{tab:LiMg_bulk}.
Since Li and Mg are simple metals, in which the electrons are nearly free electron gas, Lindhard-based KEDFs, such as WGC and WT KEDFs, are suitable to deal with them.
As expected, the MAREs of energies obtained by WGC and WT KEDFs are on the order of 0.01 for both Li and Mg systems.
In particular, the results obtained by the \TKK{m}{16} KEDF are close to those obtained by KSDFT, and the MAREs are also comparable to those obtained by WGC and WT KEDFs.
Notably, the \TKK{m}{16} KEDF is able to reproduce the slight energy difference between different configurations, such as the energy difference of 0.007 eV/atom between the fcc and hcp Mg structures, which is close to the value of 0.011 eV/atom obtained by KSDFT.
Overall, these findings highlight the excellent transferability of \TKK{m}{16} KEDF for simple metals.
Additionally, \TKK{m}{16} KEDF also shows good transferability for Mg-Al alloys, and one can refer to the Supporting Information.
}

\mc{
It would be very interesting to study transition metals using orbital-free DFT. However, only a few works have tried to tackle transition metals and two challenges still remain. First, the framework of OFDFT can hardly handle the localized $d$ electrons and new methods are needed. For example, the electron density decomposition method was proposed to examine Cu and Ag.~\cite{12B-Huang-density-decomp} In addition, angular-momentum-dependent OFDFT was proposed to investigate Ti.~\cite{13L-Ke-amd, 14B-Ke-amd} Second, well-tested local pseudopotentials for transition metals are still needed.}

% To test the transferability of the TKK KEDFs, we apply the \TKK{m}{16} KEDF to study the bulk properties of simple metal systems such as Mg and Li. 
% %
% We choose the bcc, fcc, sc, and CD crystal structures of Li, as well as the fcc, bcc, sc, and hcp crystal structures of Mg. 
% %
% The results match well with the WT and WGC KEDF results, demonstrating the excellent transferability of the \TKK{m}{16} KEDF. More details are provided in Table S3 of SI.

\subsection{\mc{Silicon}}

\begin{table*}[htbp]
	\centering
	\caption{
 Bulk properties of the cubic diamond (CD) and $\beta$-tin crystal structures of Si, i.e., the bulk modulus ($B$ in $\tx{GPa}$), the equilibrium volume ($V_0$ in $\tx{\AA}^3$/atom), and the energy of a given system ($E_0$ in eV/atom), as well as available experimental data. The MARE, as defined in Eq.~\ref{eq.mare}, is given by comparing OFDFT to KS-BLPS results.
We set the $E_0$ to be the total energy for the CD structure, while the value for the $\beta$-tin is set to the energy difference with respect to the total energy of CD Si.
Both KSDFT and OFDFT calculations are performed with the use of bulk-derived pseudopotentials (BLPS). 
For OFDFT calculations, we use the WGC and HC KEDFs, as well as the TKKs for semiconductors with different cutoffs $\lambda_c$ (labeled as TKK$^{s}_{\lambda_c}$).
}
	\begin{tabularx}{0.9\linewidth}{
			>{\raggedright\arraybackslash\hsize=0.5\hsize\linewidth=\hsize}X
			>{\raggedright\arraybackslash\hsize=1.5\hsize\linewidth=\hsize}X
			>{\centering\arraybackslash}X
			>{\centering\arraybackslash}X
			>{\centering\arraybackslash}X}
		\hline\hline
		&       &CD	    &$\beta$-tin    &MARE\\
		\hline
		$B\ (\tx{GPa})$	&KS-BLPS (this work)	&\mc{99}	&\mc{123}  &-\\
		&KS-BLPS (Ref.~\onlinecite{08PCCP-Huang-BLPS})    &\mc{98}   &\mc{122}  &-\\
		&HC~(Ref.~\onlinecite{10B-Huang-nonlocal})    &\mc{97} &\mc{83} &17.18$\%$\\
		% &HC	    &103.4	&92.8   &14.83\\
		&WGC	&\mc{54}	&\mc{140}  &29.40$\%$\\
		% &WT	    &-	    &101.4\\
		% &truncated WT	&-	    &69.7\\
		%&\TKK{s}{20}	&87.5	&105.1  &13.04\\
		&\TKK{s}{16}	&\mc{78}	&\mc{110}  &15.86$\%$\\
		&\TKK{s}{12}	&\mc{74}	&\mc{87}   &27.25$\%$\\
		&\TKK{s}{8}	    &\mc{93}	&\mc{132}  &6.45$\%$\\
		&Exp. (Ref.~\onlinecite{06-Martienssen-si_bulk_exp}) &98.0\\
		\\
		$V_0\ (\tx{\AA}^3)$ &KS-BLPS (this work)	&19.774	&14.621 &-\\
		&KS-BLPS (Ref.~\onlinecite{08PCCP-Huang-BLPS})    &19.777 &14.663 &-\\
		&HC~(Ref.~\onlinecite{10B-Huang-nonlocal})    &19.962 &15.662 &4.04$\%$\\
		% &HC	    &19.986	&14.926 &1.58\\
		&WGC	&21.504	&14.406 &5.11$\%$\\
		% &WT	    &-	    &14.660\\
		% &truncated WT	&-	    &15.706\\
		%	&\TKK{s}{20}	&19.197	&14.464 &2.00\\
		&\TKK{s}{16}	&19.470	&14.482 &1.24$\%$\\
		&\TKK{s}{12}	&19.540	&14.860 &1.41$\%$\\
		&\TKK{s}{8}	&18.974	&14.133 &3.69$\%$\\
		&Exp. (Ref.~\onlinecite{06-Martienssen-si_bulk_exp}) &20.013\\
		\\
		$E_0\ (\tx{eV})$	&KS-BLPS (this work)	&$-109.629$	&0.168  &-\\
		&KS-BLPS (Ref.~\onlinecite{08PCCP-Huang-BLPS})    &$-109.629$ &0.166  &-\\
		&HC~(Ref.~\onlinecite{10B-Huang-nonlocal})    &$-109.624$ &0.170  &0.01$\%$\\
		% 	&HC	    &$-109.610$	&0.210  &0.04\\
		&WGC	&$-109.332$	&0.016  &0.2$\%$\\
		% 	&WT	    &-	        &$-109.36$\\
		% &truncated WT	&-	        &$-109.27$\\
		%	&\TKK{s}{20}	&$-109.638$	&0.273  &0.05\\
		&\TKK{s}{16}	&$-109.583$	&0.165  &0.04$\%$\\
		&\TKK{s}{12}	&$-109.562$	&0.100  &0.03$\%$\\
		&\TKK{s}{8}	&$-109.545$	&0.035  &0.06$\%$\\
		\hline\hline
	\end{tabularx}
	\label{tab:Si_bulk}
\end{table*}

Table~\ref{tab:Si_bulk} lists the bulk properties of CD and $\beta$-tin crystal structures of Si as obtained by KSDFT and OFDFT with various KEDFs, where the CD Si is a typical semiconductor.
For the bulk modulus of CD Si, the KS-BLPS calculations yield a value of around 99 GPa, which is in excellent agreement with the experimental data of 98.0 GPa. 
Meanwhile, the KS-BLPS predicts the equilibrium volume to be 19.774 $\tx{\AA}^3$/atom, which is $1.19\%$ smaller than the experimental value of 20.013 $\tx{\AA}^3$/atom. Based on the data, we conclude that the KS-BLPS results are reasonable. 

For the OFDFT calculations, since the WT KEDF is designed for free-electron-like systems, it is not surprising that the WT KEDF
fails in calculating the bulk properties of the CD Si structure. Therefore, no WT results are included in Table~\ref{tab:Si_bulk}.
In this regard, we utilize the more sophisticated WGC KEDF, which yields a substantially smaller bulk modulus of 54 GPa as compared to the experimental value of 98.0 GPa. Worse still, the equilibrium volume from WGC is 21.504 $\tx{\AA}^3$/atom, which is 7.45$\%$ larger than the experimental value.
In addition, we list the results of the HC KEDF taken from Ref.~\onlinecite{10B-Huang-nonlocal}. The HC is designed for semiconductors and performs significantly better than the WGC KEDF for the tested properties of CD Si. Unfortunately, the HC KEDF yields worse bulk modulus and equilibrium volume for the $\beta$-tin structure when compared to WGC.

Notably, the energy difference between the CD and $\beta$-tin Si obtained from the HC KEDF (0.170 eV/atom) is close to the one from KS-BLPS (0.168 eV/atom), while the WGC KEDF yields a much smaller one of 0.016 eV/atom.
By utilizing the TKK$^{\rm{s}}$s with the increase of the cutoff from 8 to 16, we observe that the energy difference between CD and $\beta$-tin Si changes from 0.035 to 0.165 eV/atom, approaching the results of KSDFT (0.168 eV/atom).
Although the \TKK{s}{16} KEDF yields a worse bulk modulus of CD Si (78 GPa) than the HC KEDF (97 GPa), it performs better (110 GPa) than the HC (83 GPa) for the $\beta$-tin Si. Besides, the equilibrium volume of  the $\beta$-tin Si structure is 14.482 and 15.662 \AA$^3$/atom from the \TKK{s}{16} KEDF and the HC KEDF, respectively; the former one is substantially closer to the 14.621 \AA$^3$/atom as obtained from KS-BLPS calculations in this work.
Regarding the new KEDFs proposed in this work, we observe that all of the three TKK$^{\rm{s}}_{\lambda_c}$s 
perform better than the WGC KEDF, which is evidenced by the substantially smaller MARE in all three properties, including the bulk modulus, the equilibrium volume, and the total energy.
This demonstrates that the \TKK{s}{16} KEDF exhibits a better balance to describe the two phases of Si than the HC KEDF, and the long-range part of TKK plays a crucial role in determining the accuracy of non-local KEDF.

The surface energies and vacancy formation energies of Si are shown in Table~\ref{tab:Surface_vacancy}.
Since the Si systems own more localized electrons than the Al systems, and the corresponding surfaces involve the presence of a vacuum, it is not surprising that the Lindhard-based KEDFs cannot yield reasonable results for surfaces or vacancies of Si, as previous works have demonstrated this.
For example, we encounter convergence issues with the WGC KEDF when dealing with the surface and vacancy of CD Si.
Worse still, we find that the WT and the \TKK{s}{8} KEDFs predict negative surface energies and vacancy formation energies, which are qualitatively incorrect values as compared to the KSDFT data.
Interestingly, the \TKK{s}{16} KEDF yields close values as compared to the KS-BLPS method. In detail,
the \TKK{s}{16} (KS-BLPS) predicts the CD Si (100) surface energy and the vacancy formation energy (in a 2$\times$2$\times$2 cell) in CD Si phase as 2228 (2062) $\rm{mJ/m^2}$ and 3.583 (3.240) eV, respectively.
The \TKK{s}{16} KEDF performs substantially better than the HC KEDF, the latter of which predicts the vacancy formation energy as 1.445 eV.
In addition, \TKK{s}{12} also yields a reasonable vacancy formation energy of 3.143 eV but a lower surface energy of 1307 $\rm{mJ/m^2}$.
Therefore, we conclude that the \TKK{s}{16} performs better than the WT, WGC, and HC KEDFs for the surface energy and vacancy formation energy of Si. 
We plot the convergence trend of vacancy energies with respect to system size in Fig.~S2 of SI. We notice that in systems containing over one hundred atoms, the results obtained through the \TKK{s}{16} KEDF are almost indistinguishable from those obtained through KSDFT. The results again demonstrate the excellent performance of the \TKK{s}{16} KEDF.

We further compare the electron density differences of bulk CD Si and its (100) surface as obtained from OFDFT and KSDFT calculations, which are displayed in Fig.~\ref{fig:cdSi_den}. Note that we choose the (010) crystal surface of a bulk Si configuration to represent the electron density differences of bulk Si.
The representative electron density profiles of the bulk Si and the $(1\bar{1}0)$ surface of Si are shown in Figs.~\ref{fig:cdSi_den}(a) and (c), respectively.
As shown in Fig.~\ref{fig:cdSi_den}(b), the MAREs of electron density in bulk CD Si as obtained from the HC, WT, and \TKK{s}{16} KEDFs are 0.07, 0.12, 0.10, respectively.
The results are two orders of magnitude larger than those in bulk fcc Al, indicating that the electronic structure of semiconductors is more challenging to describe by KEDFs than the metallic ones.
Fig.~\ref{fig:cdSi_den}(d) illustrates the electron density differences on the longitudinal section of the Si CD (100) surface, as well as the MAREs.
\mc{As explained before, small electron density in vacuum may result in a large contribution to the MARE defined in Eq.~\ref{eq:mare}, so we did not calculate the contributions to MARE for a selected length of vacuum (6.0 Å) in the Si slab. As a result, we find the MAREs obtained by the HC, WT, and \TKK{s}{16} KEDFs are 0.18, 0.14, and 0.16, respectively. We notice that the MAREs of surface system and bulk system are on the same order of magnitude. 
The MAREs of Si CD (100) surface are slightly larger than those of bulk CD Si. This can be explained by the fact that most KEDFs are not suitable for describing the covalent bonds of Si formed by electrons.}
\mc{From Fig.~\ref{fig:cdSi_den}(d), we also observe that the WT and \TKK{s}{16} KEDFs share a similar pattern of electron density differences. In future, one can test the revHC KEDF~\cite{21B-Shao-nonlocal} and see its performance for the above tests.}
%
% The results suggest that the HC KEDF yields substantially worse electron density for the Si (100) surface. 
Although the WT KEDF gives a better electron density, it predicts qualitatively wrong surface energy and vacancy formation energies, as listed in Table~\ref{tab:Surface_vacancy}. Among the three KEDFs, \mc{both HC and \TKK{s}{16} KEDFs are} capable of capturing both the energies and the electron density with similar accuracy to KSDFT.

\section{Conclusion}

In this work, we constructed two groups of TKKs with different cutoffs for metals and semiconductor systems, \mc{as a first step to find an optimal KEDF for metals and semiconductors}. We further compared the performances of these kernels to validate how the real-space cutoff affects the properties of Al and Si systems.
We systematically investigated several properties of the bulk and surface structures of Al and Si.

In general, the accuracy of TKKs increases with a larger cutoff. However, we found the TKK KEDFs with a short-ranged kinetic energy kernel ($\lambda_c =8,12$) yielded unreasonable stacking fault energies, surface energies, and vacancy formation energies for Al systems. Interestingly, we found that when the real-space cutoff of the TKK was larger than the distance between an atom and its next nearest neighbor atoms, the TKK was able to accurately characterize these properties and performed even better than the WT KEDF.
In conclusion, considering the interactions between an atom and its next nearest neighbor atoms is crucial for a non-local KEDF to distinguish the energy orderings among bulk structures, such as the fcc and hcp solid phases of Al, and CD and $\beta$-tin solid phases of Si. Furthermore, it helps to accurately predict the surface energies and point vacancies of Al and Si systems.

We found the \TKK{m}{16} and \TKK{s}{16} kernels presented in this work gave reasonable results in all of the above tests. In addition, these kernels even performed better than the WT, WGC, and HC KEDFs in some aspects.
For example, \TKK{m}{16} yielded more accurate stacking fault energies than the WGC and WT KEDFs for the fcc structure of Al. The kernel performed better than the WT KEDF when dealing with the surface and vacancy formation energies in fcc Al.
On the other hand, the \TKK{s}{16} kernel yielded better vacancy formation energies than the HC KEDF for the CD Si structure. It exhibited reasonable accuracy in predicting the electron densities for bulk Al and Si systems, as well as the Al fcc (100) and Si CD (100) surfaces.

Despite the above advantages of the newly proposed TKKs, we also encountered issues \mc{in the following three aspects.
First, \TKK{m}{}s, which are designed for metals, are not suitable for semiconductor systems such as Si.
On the other hand, \TKK{s}{}s (designed for semiconductors) are not accurate for metallic systems such as Al.
This may be attributed to the different asymptotic behavior of KEDFs for semiconductors and metals~\cite{10B-Huang-nonlocal}, and we consider that machine learning is a potential tool to achieve a global KEDF for both metals and semiconductors.}
% Second, the electronic densities of Si CD(100) surface obtained by \TKK{s}{}s share the same issue with the usage of the WT KEDF, i.e., the electrons tend to gather around the outer atoms instead of the inner atoms.
% This suggests that both WT KEDF and \TKK{s}{} are not able to well describe the covalent bonds, especially near the surfaces.
Second, the discrepancies between the electron densities obtained by OFDFT and KSDFT for semiconductors are considerably larger when compared to those observed in metals.
\mc{In particular, for the CD phase of silicon, it is still challenging to pose a truncated KEDF kernel that share the same accuracy as KSDFT. In future, it would be interesting to test more solid phases of Si.
}
\mc{Third}, as expected, these differences tend to be substantially larger in surface systems than in bulk systems.
However, we note that all of the WGC, HC, WT, and TKK KEDFs suffer from the above issues.

\mc{The force calculations with the usage of the TKK kernels have been implemented, and we found the current TKK KEDFs can be used to relax the bulk structure or even perform molecular dynamics simulations. However, the surface structures relaxed by the TKK KEDFs still deviate from the KSDFT results. One of the reasons is that the forces were not included in the residual function, and we expect the geometry relaxation and molecular dynamics functions be tested in future works.}
%

%In conclusion, we explore the possibility of using non-local KEDFs to describe metal and semiconductor systems, which deepens our understanding of the forms and accuracy of KEDFs and sheds new light on designing new forms of KEDFs.

To sum up, our investigation into the feasibility of employing non-local KEDFs in characterizing \mc{simple metal and Si} systems enhances our comprehension of the forms and precision of KEDFs. 
Additionally, it sheds new light on designing novel KEDFs. \mc{For future studies of other systems, such as molecules, insulators, and transition metals, etc. Two challenges should be overcome. First, the generation of transferable local pseudopotentials.~\cite{04B-Zhou-pseudo, 08PCCP-Huang-BLPS, 15CPL-Legrain-pesudo, 17JCTC-Del-pseudo} Second, more general KEDFs that can be applied to a variety of systems are needed.}

% \section*{Acknowledgements}
\acknowledgements

The work of L.S., Y.L., and M.C. was supported by the National Science Foundation of China under Grand No. 12074007 and No. 12122401. The numerical simulations were performed on the High-Performance Computing Platform of CAPT and the Bohrium platform supported by DP Technology.

\bibliography{OF-KEDF}

\begin{thebibliography}{61}
\expandafter\ifx\csname natexlab\endcsname\relax\def\natexlab#1{#1}\fi
\expandafter\ifx\csname bibnamefont\endcsname\relax
  \def\bibnamefont#1{#1}\fi
\expandafter\ifx\csname bibfnamefont\endcsname\relax
  \def\bibfnamefont#1{#1}\fi
\expandafter\ifx\csname citenamefont\endcsname\relax
  \def\citenamefont#1{#1}\fi
\expandafter\ifx\csname url\endcsname\relax
  \def\url#1{\texttt{#1}}\fi
\expandafter\ifx\csname urlprefix\endcsname\relax\def\urlprefix{URL }\fi
\providecommand{\bibinfo}[2]{#2}
\providecommand{\eprint}[2][]{\url{#2}}

\bibitem[{\citenamefont{Hohenberg and Kohn}(1964)}]{64PR-Hohenberg}
\bibinfo{author}{\bibfnamefont{P.}~\bibnamefont{Hohenberg}} \bibnamefont{and}
  \bibinfo{author}{\bibfnamefont{W.}~\bibnamefont{Kohn}},
  \bibinfo{journal}{Phys. Rev.} \textbf{\bibinfo{volume}{136}},
  \bibinfo{pages}{864B} (\bibinfo{year}{1964}).

\bibitem[{\citenamefont{Kohn and Sham}(1965)}]{65PR-Kohn}
\bibinfo{author}{\bibfnamefont{W.}~\bibnamefont{Kohn}} \bibnamefont{and}
  \bibinfo{author}{\bibfnamefont{L.~J.} \bibnamefont{Sham}},
  \bibinfo{journal}{Phys. Rev.} \textbf{\bibinfo{volume}{140}},
  \bibinfo{pages}{1133A} (\bibinfo{year}{1965}).

\bibitem[{\citenamefont{Wang and Carter}(2002)}]{02Carter}
\bibinfo{author}{\bibfnamefont{Y.~A.} \bibnamefont{Wang}} \bibnamefont{and}
  \bibinfo{author}{\bibfnamefont{E.~A.} \bibnamefont{Carter}},
  \bibinfo{journal}{Theoretical Methods in Condensed Phase Chemistry} p.
  \bibinfo{pages}{117} (\bibinfo{year}{2002}).

\bibitem[{\citenamefont{Witt et~al.}(2018)\citenamefont{Witt, Beatriz,
  Dieterich, and Carter}}]{18JMR-Witt}
\bibinfo{author}{\bibfnamefont{W.~C.} \bibnamefont{Witt}},
  \bibinfo{author}{\bibfnamefont{G.}~\bibnamefont{Beatriz}},
  \bibinfo{author}{\bibfnamefont{J.~M.} \bibnamefont{Dieterich}},
  \bibnamefont{and} \bibinfo{author}{\bibfnamefont{E.~A.}
  \bibnamefont{Carter}}, \bibinfo{journal}{J. Mater. Res.}
  \textbf{\bibinfo{volume}{33}}, \bibinfo{pages}{777} (\bibinfo{year}{2018}).

\bibitem[{\citenamefont{Shin and Carter}(2014{\natexlab{a}})}]{14AM-Shin}
\bibinfo{author}{\bibfnamefont{I.}~\bibnamefont{Shin}} \bibnamefont{and}
  \bibinfo{author}{\bibfnamefont{E.~A.} \bibnamefont{Carter}},
  \bibinfo{journal}{Acta Mater.} \textbf{\bibinfo{volume}{64}},
  \bibinfo{pages}{198} (\bibinfo{year}{2014}{\natexlab{a}}).

\bibitem[{\citenamefont{Zhuang et~al.}(2017)\citenamefont{Zhuang, Chen, and
  Carter}}]{17MSMSE-Zhuang}
\bibinfo{author}{\bibfnamefont{H.~L.} \bibnamefont{Zhuang}},
  \bibinfo{author}{\bibfnamefont{M.}~\bibnamefont{Chen}}, \bibnamefont{and}
  \bibinfo{author}{\bibfnamefont{E.~A.} \bibnamefont{Carter}},
  \bibinfo{journal}{Model. Simul. Mater. Sci. Eng.}
  \textbf{\bibinfo{volume}{25}}, \bibinfo{pages}{075002}
  (\bibinfo{year}{2017}).

\bibitem[{\citenamefont{Zhuang et~al.}(2018)\citenamefont{Zhuang, Chen, and
  Carter}}]{18PRM-Zhuang}
\bibinfo{author}{\bibfnamefont{H.~L.} \bibnamefont{Zhuang}},
  \bibinfo{author}{\bibfnamefont{M.}~\bibnamefont{Chen}}, \bibnamefont{and}
  \bibinfo{author}{\bibfnamefont{E.~A.} \bibnamefont{Carter}},
  \bibinfo{journal}{Phys. Rev. Mater.} \textbf{\bibinfo{volume}{2}},
  \bibinfo{pages}{073603} (\bibinfo{year}{2018}).

\bibitem[{\citenamefont{Witt et~al.}(2021)\citenamefont{Witt, Shires, Tan,
  Jankowski, and Pickard}}]{21JPCA-Witt}
\bibinfo{author}{\bibfnamefont{W.~C.} \bibnamefont{Witt}},
  \bibinfo{author}{\bibfnamefont{B.~W.} \bibnamefont{Shires}},
  \bibinfo{author}{\bibfnamefont{C.~W.} \bibnamefont{Tan}},
  \bibinfo{author}{\bibfnamefont{W.~J.} \bibnamefont{Jankowski}},
  \bibnamefont{and} \bibinfo{author}{\bibfnamefont{C.~J.}
  \bibnamefont{Pickard}}, \bibinfo{journal}{J. Phys. Chem. A}
  \textbf{\bibinfo{volume}{125}}, \bibinfo{pages}{1650} (\bibinfo{year}{2021}).

\bibitem[{\citenamefont{Chen et~al.}(2013)\citenamefont{Chen, Hung, Huang, Xia,
  and Carter}}]{13MP-Chen}
\bibinfo{author}{\bibfnamefont{M.}~\bibnamefont{Chen}},
  \bibinfo{author}{\bibfnamefont{L.}~\bibnamefont{Hung}},
  \bibinfo{author}{\bibfnamefont{C.}~\bibnamefont{Huang}},
  \bibinfo{author}{\bibfnamefont{J.}~\bibnamefont{Xia}}, \bibnamefont{and}
  \bibinfo{author}{\bibfnamefont{E.~A.} \bibnamefont{Carter}},
  \bibinfo{journal}{Mol. Phys.} \textbf{\bibinfo{volume}{111}},
  \bibinfo{pages}{3448} (\bibinfo{year}{2013}).

\bibitem[{\citenamefont{Mi and Pavanello}(2019)}]{19B-Mi-qdot}
\bibinfo{author}{\bibfnamefont{W.}~\bibnamefont{Mi}} \bibnamefont{and}
  \bibinfo{author}{\bibfnamefont{M.}~\bibnamefont{Pavanello}},
  \bibinfo{journal}{Phys. Rev. B} \textbf{\bibinfo{volume}{100}},
  \bibinfo{pages}{041105} (\bibinfo{year}{2019}).

\bibitem[{\citenamefont{Xu et~al.}(2020)\citenamefont{Xu, Lv, Wang, and
  Ma}}]{20B-Xu-nonlocal}
\bibinfo{author}{\bibfnamefont{Q.}~\bibnamefont{Xu}},
  \bibinfo{author}{\bibfnamefont{J.}~\bibnamefont{Lv}},
  \bibinfo{author}{\bibfnamefont{Y.}~\bibnamefont{Wang}}, \bibnamefont{and}
  \bibinfo{author}{\bibfnamefont{Y.}~\bibnamefont{Ma}}, \bibinfo{journal}{Phys.
  Rev. B} \textbf{\bibinfo{volume}{101}}, \bibinfo{pages}{045110}
  (\bibinfo{year}{2020}).

\bibitem[{\citenamefont{Liu et~al.}(2020)\citenamefont{Liu, Lu, and
  Chen}}]{20JPCM-QianruiLiu-wdm}
\bibinfo{author}{\bibfnamefont{Q.}~\bibnamefont{Liu}},
  \bibinfo{author}{\bibfnamefont{D.}~\bibnamefont{Lu}}, \bibnamefont{and}
  \bibinfo{author}{\bibfnamefont{M.}~\bibnamefont{Chen}}, \bibinfo{journal}{J.
  Phys.: Condens. Matter} \textbf{\bibinfo{volume}{32}},
  \bibinfo{pages}{144002} (\bibinfo{year}{2020}).

\bibitem[{\citenamefont{Kang et~al.}(2020)\citenamefont{Kang, Luo, Runge, and
  Trickey}}]{20Kang-semilocal}
\bibinfo{author}{\bibfnamefont{D.}~\bibnamefont{Kang}},
  \bibinfo{author}{\bibfnamefont{K.}~\bibnamefont{Luo}},
  \bibinfo{author}{\bibfnamefont{K.}~\bibnamefont{Runge}}, \bibnamefont{and}
  \bibinfo{author}{\bibfnamefont{S.}~\bibnamefont{Trickey}},
  \bibinfo{journal}{Matter Radiat. at Extremes} \textbf{\bibinfo{volume}{5}},
  \bibinfo{pages}{064403} (\bibinfo{year}{2020}).

\bibitem[{\citenamefont{Ding et~al.}(2018)\citenamefont{Ding, White, Hu,
  Certik, and Collins}}]{18L-Ding-oftddft}
\bibinfo{author}{\bibfnamefont{Y.}~\bibnamefont{Ding}},
  \bibinfo{author}{\bibfnamefont{A.~J.} \bibnamefont{White}},
  \bibinfo{author}{\bibfnamefont{S.}~\bibnamefont{Hu}},
  \bibinfo{author}{\bibfnamefont{O.}~\bibnamefont{Certik}}, \bibnamefont{and}
  \bibinfo{author}{\bibfnamefont{L.~A.} \bibnamefont{Collins}},
  \bibinfo{journal}{Phys. Rev. Lett.} \textbf{\bibinfo{volume}{121}},
  \bibinfo{pages}{145001} (\bibinfo{year}{2018}).

\bibitem[{\citenamefont{White et~al.}(2018)\citenamefont{White, Certik, Ding,
  Hu, and Collins}}]{18B-White-oftddft}
\bibinfo{author}{\bibfnamefont{A.~J.} \bibnamefont{White}},
  \bibinfo{author}{\bibfnamefont{O.}~\bibnamefont{Certik}},
  \bibinfo{author}{\bibfnamefont{Y.}~\bibnamefont{Ding}},
  \bibinfo{author}{\bibfnamefont{S.}~\bibnamefont{Hu}}, \bibnamefont{and}
  \bibinfo{author}{\bibfnamefont{L.~A.} \bibnamefont{Collins}},
  \bibinfo{journal}{Phys. Rev. B} \textbf{\bibinfo{volume}{98}},
  \bibinfo{pages}{144302} (\bibinfo{year}{2018}).

\bibitem[{\citenamefont{Xiang et~al.}(2019)\citenamefont{Xiang, Wang, Xu,
  Zhang, and Lu}}]{20JPC-Xiang-oftddft}
\bibinfo{author}{\bibfnamefont{H.}~\bibnamefont{Xiang}},
  \bibinfo{author}{\bibfnamefont{Z.}~\bibnamefont{Wang}},
  \bibinfo{author}{\bibfnamefont{L.}~\bibnamefont{Xu}},
  \bibinfo{author}{\bibfnamefont{X.}~\bibnamefont{Zhang}}, \bibnamefont{and}
  \bibinfo{author}{\bibfnamefont{G.}~\bibnamefont{Lu}}, \bibinfo{journal}{J.
  Phys. Chem. C} \textbf{\bibinfo{volume}{124}}, \bibinfo{pages}{945}
  (\bibinfo{year}{2019}).

\bibitem[{\citenamefont{Jiang et~al.}(2021)\citenamefont{Jiang, Shao, Pavanello
  et~al.}}]{21B-Jiang-oftddft}
\bibinfo{author}{\bibfnamefont{K.}~\bibnamefont{Jiang}},
  \bibinfo{author}{\bibfnamefont{X.}~\bibnamefont{Shao}},
  \bibinfo{author}{\bibfnamefont{M.}~\bibnamefont{Pavanello}},
  \bibnamefont{et~al.}, \bibinfo{journal}{Phys. Rev. B}
  \textbf{\bibinfo{volume}{104}}, \bibinfo{pages}{235110}
  (\bibinfo{year}{2021}).

\bibitem[{\citenamefont{Constantin
  et~al.}(2018{\natexlab{a}})\citenamefont{Constantin, Fabiano, and
  Della~Sala}}]{18JPCL-Constantin-semilocal}
\bibinfo{author}{\bibfnamefont{L.~A.} \bibnamefont{Constantin}},
  \bibinfo{author}{\bibfnamefont{E.}~\bibnamefont{Fabiano}}, \bibnamefont{and}
  \bibinfo{author}{\bibfnamefont{F.}~\bibnamefont{Della~Sala}},
  \bibinfo{journal}{J. Phys. Chem. Lett.} \textbf{\bibinfo{volume}{9}},
  \bibinfo{pages}{4385} (\bibinfo{year}{2018}{\natexlab{a}}).

\bibitem[{\citenamefont{Constantin et~al.}(2019)\citenamefont{Constantin,
  Fabiano, and Della~Sala}}]{19JCTC-Constantin-semilocal}
\bibinfo{author}{\bibfnamefont{L.~A.} \bibnamefont{Constantin}},
  \bibinfo{author}{\bibfnamefont{E.}~\bibnamefont{Fabiano}}, \bibnamefont{and}
  \bibinfo{author}{\bibfnamefont{F.}~\bibnamefont{Della~Sala}},
  \bibinfo{journal}{J. Chem. Theory Comput.} \textbf{\bibinfo{volume}{15}},
  \bibinfo{pages}{3044} (\bibinfo{year}{2019}).

\bibitem[{\citenamefont{Luo et~al.}(2018)\citenamefont{Luo, Karasiev, and
  Trickey}}]{18B-Luo-semilocal}
\bibinfo{author}{\bibfnamefont{K.}~\bibnamefont{Luo}},
  \bibinfo{author}{\bibfnamefont{V.~V.} \bibnamefont{Karasiev}},
  \bibnamefont{and} \bibinfo{author}{\bibfnamefont{S.}~\bibnamefont{Trickey}},
  \bibinfo{journal}{Phys. Rev. B} \textbf{\bibinfo{volume}{98}},
  \bibinfo{pages}{041111} (\bibinfo{year}{2018}).

\bibitem[{\citenamefont{Luo et~al.}(2020)\citenamefont{Luo, Karasiev, and
  Trickey}}]{20B-Luo-semilocal}
\bibinfo{author}{\bibfnamefont{K.}~\bibnamefont{Luo}},
  \bibinfo{author}{\bibfnamefont{V.~V.} \bibnamefont{Karasiev}},
  \bibnamefont{and} \bibinfo{author}{\bibfnamefont{S.}~\bibnamefont{Trickey}},
  \bibinfo{journal}{Phys. Rev. B} \textbf{\bibinfo{volume}{101}},
  \bibinfo{pages}{075116} (\bibinfo{year}{2020}).

\bibitem[{\citenamefont{Wang and Teter}(1992)}]{92B-Wang-nonlocal}
\bibinfo{author}{\bibfnamefont{L.-W.} \bibnamefont{Wang}} \bibnamefont{and}
  \bibinfo{author}{\bibfnamefont{M.~P.} \bibnamefont{Teter}},
  \bibinfo{journal}{Phys. Rev. B} \textbf{\bibinfo{volume}{45}},
  \bibinfo{pages}{13196} (\bibinfo{year}{1992}).

\bibitem[{\citenamefont{Smargiassi and Madden}(1994)}]{94B-Smargiassi-nonlocal}
\bibinfo{author}{\bibfnamefont{E.}~\bibnamefont{Smargiassi}} \bibnamefont{and}
  \bibinfo{author}{\bibfnamefont{P.~A.} \bibnamefont{Madden}},
  \bibinfo{journal}{Phys. Rev. B} \textbf{\bibinfo{volume}{49}},
  \bibinfo{pages}{5220} (\bibinfo{year}{1994}).

\bibitem[{\citenamefont{Wang et~al.}(1999)\citenamefont{Wang, Govind, and
  Carter}}]{99B-Wang-nonlocal}
\bibinfo{author}{\bibfnamefont{Y.~A.} \bibnamefont{Wang}},
  \bibinfo{author}{\bibfnamefont{N.}~\bibnamefont{Govind}}, \bibnamefont{and}
  \bibinfo{author}{\bibfnamefont{E.~A.} \bibnamefont{Carter}},
  \bibinfo{journal}{Phys. Rev. B} \textbf{\bibinfo{volume}{60}},
  \bibinfo{pages}{16350} (\bibinfo{year}{1999}).

\bibitem[{\citenamefont{Huang and Carter}(2010)}]{10B-Huang-nonlocal}
\bibinfo{author}{\bibfnamefont{C.}~\bibnamefont{Huang}} \bibnamefont{and}
  \bibinfo{author}{\bibfnamefont{E.~A.} \bibnamefont{Carter}},
  \bibinfo{journal}{Phys. Rev. B} \textbf{\bibinfo{volume}{81}},
  \bibinfo{pages}{045206} (\bibinfo{year}{2010}).

\bibitem[{\citenamefont{Shin and
  Carter}(2014{\natexlab{b}})}]{14JCP-Shin-nonlocal}
\bibinfo{author}{\bibfnamefont{I.}~\bibnamefont{Shin}} \bibnamefont{and}
  \bibinfo{author}{\bibfnamefont{E.~A.} \bibnamefont{Carter}},
  \bibinfo{journal}{J. Chem. Phys.} \textbf{\bibinfo{volume}{140}},
  \bibinfo{pages}{18A531} (\bibinfo{year}{2014}{\natexlab{b}}).

\bibitem[{\citenamefont{Constantin
  et~al.}(2018{\natexlab{b}})\citenamefont{Constantin, Fabiano, and
  Della~Sala}}]{18B-Constantin-nonlocal}
\bibinfo{author}{\bibfnamefont{L.~A.} \bibnamefont{Constantin}},
  \bibinfo{author}{\bibfnamefont{E.}~\bibnamefont{Fabiano}}, \bibnamefont{and}
  \bibinfo{author}{\bibfnamefont{F.}~\bibnamefont{Della~Sala}},
  \bibinfo{journal}{Phys. Rev. B} \textbf{\bibinfo{volume}{97}},
  \bibinfo{pages}{205137} (\bibinfo{year}{2018}{\natexlab{b}}).

\bibitem[{\citenamefont{Mi et~al.}(2018)\citenamefont{Mi, Genova, and
  Pavanello}}]{18JCP-Mi-nonlocal}
\bibinfo{author}{\bibfnamefont{W.}~\bibnamefont{Mi}},
  \bibinfo{author}{\bibfnamefont{A.}~\bibnamefont{Genova}}, \bibnamefont{and}
  \bibinfo{author}{\bibfnamefont{M.}~\bibnamefont{Pavanello}},
  \bibinfo{journal}{J. Chem. Phys.} \textbf{\bibinfo{volume}{148}},
  \bibinfo{pages}{184107} (\bibinfo{year}{2018}).

\bibitem[{\citenamefont{Xu et~al.}(2019)\citenamefont{Xu, Wang, and
  Ma}}]{19B-Xu-nonlocal}
\bibinfo{author}{\bibfnamefont{Q.}~\bibnamefont{Xu}},
  \bibinfo{author}{\bibfnamefont{Y.}~\bibnamefont{Wang}}, \bibnamefont{and}
  \bibinfo{author}{\bibfnamefont{Y.}~\bibnamefont{Ma}}, \bibinfo{journal}{Phys.
  Rev. B} \textbf{\bibinfo{volume}{100}}, \bibinfo{pages}{205132}
  (\bibinfo{year}{2019}).

\bibitem[{\citenamefont{Shao et~al.}(2021)\citenamefont{Shao, Mi, and
  Pavanello}}]{21B-Shao-nonlocal}
\bibinfo{author}{\bibfnamefont{X.}~\bibnamefont{Shao}},
  \bibinfo{author}{\bibfnamefont{W.}~\bibnamefont{Mi}}, \bibnamefont{and}
  \bibinfo{author}{\bibfnamefont{M.}~\bibnamefont{Pavanello}},
  \bibinfo{journal}{Phys. Rev. B} \textbf{\bibinfo{volume}{104}},
  \bibinfo{pages}{045118} (\bibinfo{year}{2021}).

\bibitem[{\citenamefont{Chen et~al.}(2016)\citenamefont{Chen, Jiang, Zhuang,
  Wang, and Carter}}]{16JCTC-Chen}
\bibinfo{author}{\bibfnamefont{M.}~\bibnamefont{Chen}},
  \bibinfo{author}{\bibfnamefont{X.-W.} \bibnamefont{Jiang}},
  \bibinfo{author}{\bibfnamefont{H.}~\bibnamefont{Zhuang}},
  \bibinfo{author}{\bibfnamefont{L.-W.} \bibnamefont{Wang}}, \bibnamefont{and}
  \bibinfo{author}{\bibfnamefont{E.~A.} \bibnamefont{Carter}},
  \bibinfo{journal}{J. Chem. Theory Comput.} \textbf{\bibinfo{volume}{12}},
  \bibinfo{pages}{2950} (\bibinfo{year}{2016}).

\bibitem[{\citenamefont{Haynes and Payne}(1997)}]{97CPC-Haynes-bessel}
\bibinfo{author}{\bibfnamefont{P.~D.} \bibnamefont{Haynes}} \bibnamefont{and}
  \bibinfo{author}{\bibfnamefont{M.~C.} \bibnamefont{Payne}},
  \bibinfo{journal}{Comp. Phys. Commun.} \textbf{\bibinfo{volume}{102}},
  \bibinfo{pages}{17} (\bibinfo{year}{1997}).

\bibitem[{\citenamefont{Chen et~al.}(2010)\citenamefont{Chen, Guo, and
  He}}]{10JPCM-Chen-bessel}
\bibinfo{author}{\bibfnamefont{M.}~\bibnamefont{Chen}},
  \bibinfo{author}{\bibfnamefont{G.}~\bibnamefont{Guo}}, \bibnamefont{and}
  \bibinfo{author}{\bibfnamefont{L.}~\bibnamefont{He}}, \bibinfo{journal}{J.
  Phys.: Condens. Matter} \textbf{\bibinfo{volume}{22}},
  \bibinfo{pages}{445501} (\bibinfo{year}{2010}).

\bibitem[{\citenamefont{Chen et~al.}(2011)\citenamefont{Chen, Guo, and
  He}}]{11JPCM-Chen-bessel}
\bibinfo{author}{\bibfnamefont{M.}~\bibnamefont{Chen}},
  \bibinfo{author}{\bibfnamefont{G.}~\bibnamefont{Guo}}, \bibnamefont{and}
  \bibinfo{author}{\bibfnamefont{L.}~\bibnamefont{He}}, \bibinfo{journal}{J.
  Phys.: Condens. Matter} \textbf{\bibinfo{volume}{23}},
  \bibinfo{pages}{325501} (\bibinfo{year}{2011}).

\bibitem[{\citenamefont{Li et~al.}(2016{\natexlab{a}})\citenamefont{Li, Liu,
  Chen, Lin, Ren, Lin, Yang, and He}}]{16CMS-Li-bessel}
\bibinfo{author}{\bibfnamefont{P.}~\bibnamefont{Li}},
  \bibinfo{author}{\bibfnamefont{X.}~\bibnamefont{Liu}},
  \bibinfo{author}{\bibfnamefont{M.}~\bibnamefont{Chen}},
  \bibinfo{author}{\bibfnamefont{P.}~\bibnamefont{Lin}},
  \bibinfo{author}{\bibfnamefont{X.}~\bibnamefont{Ren}},
  \bibinfo{author}{\bibfnamefont{L.}~\bibnamefont{Lin}},
  \bibinfo{author}{\bibfnamefont{C.}~\bibnamefont{Yang}}, \bibnamefont{and}
  \bibinfo{author}{\bibfnamefont{L.}~\bibnamefont{He}}, \bibinfo{journal}{Comp.
  Mater. Sci.} \textbf{\bibinfo{volume}{112}}, \bibinfo{pages}{503}
  (\bibinfo{year}{2016}{\natexlab{a}}).

\bibitem[{\citenamefont{Kumar et~al.}(2022)\citenamefont{Kumar, Sadigh, Zhu,
  Suryanarayana, Hamel, Gallagher, Bulatov, Klepeis, and
  Samanta}}]{22JCP-Kumar-Gaussiankernel}
\bibinfo{author}{\bibfnamefont{S.}~\bibnamefont{Kumar}},
  \bibinfo{author}{\bibfnamefont{B.}~\bibnamefont{Sadigh}},
  \bibinfo{author}{\bibfnamefont{S.}~\bibnamefont{Zhu}},
  \bibinfo{author}{\bibfnamefont{P.}~\bibnamefont{Suryanarayana}},
  \bibinfo{author}{\bibfnamefont{S.}~\bibnamefont{Hamel}},
  \bibinfo{author}{\bibfnamefont{B.}~\bibnamefont{Gallagher}},
  \bibinfo{author}{\bibfnamefont{V.}~\bibnamefont{Bulatov}},
  \bibinfo{author}{\bibfnamefont{J.}~\bibnamefont{Klepeis}}, \bibnamefont{and}
  \bibinfo{author}{\bibfnamefont{A.}~\bibnamefont{Samanta}},
  \bibinfo{journal}{J. Chem. Phys.} \textbf{\bibinfo{volume}{156}},
  \bibinfo{pages}{024107} (\bibinfo{year}{2022}).

\bibitem[{\citenamefont{Metropolis et~al.}(1953)\citenamefont{Metropolis,
  Rosenbluth, Rosenbluth, Teller, and Teller}}]{53JCP-Metropolis}
\bibinfo{author}{\bibfnamefont{N.}~\bibnamefont{Metropolis}},
  \bibinfo{author}{\bibfnamefont{A.~W.} \bibnamefont{Rosenbluth}},
  \bibinfo{author}{\bibfnamefont{M.~N.} \bibnamefont{Rosenbluth}},
  \bibinfo{author}{\bibfnamefont{A.~H.} \bibnamefont{Teller}},
  \bibnamefont{and} \bibinfo{author}{\bibfnamefont{E.}~\bibnamefont{Teller}},
  \bibinfo{journal}{J. Chem. Phys.} \textbf{\bibinfo{volume}{21}},
  \bibinfo{pages}{1087} (\bibinfo{year}{1953}).

\bibitem[{\citenamefont{Kirkpatrick et~al.}(1983)\citenamefont{Kirkpatrick,
  Gelatt~Jr, and Vecchi}}]{83Science-Kirkpatrick}
\bibinfo{author}{\bibfnamefont{S.}~\bibnamefont{Kirkpatrick}},
  \bibinfo{author}{\bibfnamefont{C.~D.} \bibnamefont{Gelatt~Jr}},
  \bibnamefont{and} \bibinfo{author}{\bibfnamefont{M.~P.}
  \bibnamefont{Vecchi}}, \bibinfo{journal}{Science}
  \textbf{\bibinfo{volume}{220}}, \bibinfo{pages}{671} (\bibinfo{year}{1983}).

\bibitem[{\citenamefont{Fermi}(1927)}]{27TANL-Fermi-local}
\bibinfo{author}{\bibfnamefont{E.}~\bibnamefont{Fermi}},
  \bibinfo{journal}{Rend. Accad. Naz. Lincei} \textbf{\bibinfo{volume}{6}},
  \bibinfo{pages}{5} (\bibinfo{year}{1927}).

\bibitem[{\citenamefont{Weizs{\"a}cker}(1935)}]{35-vW-semilocal}
\bibinfo{author}{\bibfnamefont{C.~v.} \bibnamefont{Weizs{\"a}cker}},
  \bibinfo{journal}{Zeitschrift f{\"u}r Physik} \textbf{\bibinfo{volume}{96}},
  \bibinfo{pages}{431} (\bibinfo{year}{1935}).

\bibitem[{\citenamefont{Chen et~al.}(2015)\citenamefont{Chen, Xia, Huang,
  Dieterich, Hung, Shin, and Carter}}]{15CPC-Chen-PROFESS}
\bibinfo{author}{\bibfnamefont{M.}~\bibnamefont{Chen}},
  \bibinfo{author}{\bibfnamefont{J.}~\bibnamefont{Xia}},
  \bibinfo{author}{\bibfnamefont{C.}~\bibnamefont{Huang}},
  \bibinfo{author}{\bibfnamefont{J.~M.} \bibnamefont{Dieterich}},
  \bibinfo{author}{\bibfnamefont{L.}~\bibnamefont{Hung}},
  \bibinfo{author}{\bibfnamefont{I.}~\bibnamefont{Shin}}, \bibnamefont{and}
  \bibinfo{author}{\bibfnamefont{E.~A.} \bibnamefont{Carter}},
  \bibinfo{journal}{Comput. Phys. Commun.} \textbf{\bibinfo{volume}{190}},
  \bibinfo{pages}{228} (\bibinfo{year}{2015}).

\bibitem[{\citenamefont{Li et~al.}(2016{\natexlab{b}})\citenamefont{Li, Liu,
  Chen, Lin, Ren, Lin, Yang, and He}}]{16Li-CMS-ABACUS}
\bibinfo{author}{\bibfnamefont{P.}~\bibnamefont{Li}},
  \bibinfo{author}{\bibfnamefont{X.}~\bibnamefont{Liu}},
  \bibinfo{author}{\bibfnamefont{M.}~\bibnamefont{Chen}},
  \bibinfo{author}{\bibfnamefont{P.}~\bibnamefont{Lin}},
  \bibinfo{author}{\bibfnamefont{X.}~\bibnamefont{Ren}},
  \bibinfo{author}{\bibfnamefont{L.}~\bibnamefont{Lin}},
  \bibinfo{author}{\bibfnamefont{C.}~\bibnamefont{Yang}}, \bibnamefont{and}
  \bibinfo{author}{\bibfnamefont{L.}~\bibnamefont{He}}, \bibinfo{journal}{Comp.
  Mater. Sci.} \textbf{\bibinfo{volume}{112}}, \bibinfo{pages}{503}
  (\bibinfo{year}{2016}{\natexlab{b}}).

\bibitem[{\citenamefont{Monkhorst and Pack}(1976)}]{76B-Monkhorst}
\bibinfo{author}{\bibfnamefont{H.~J.} \bibnamefont{Monkhorst}}
  \bibnamefont{and} \bibinfo{author}{\bibfnamefont{J.~D.} \bibnamefont{Pack}},
  \bibinfo{journal}{Phys. Rev. B} \textbf{\bibinfo{volume}{13}},
  \bibinfo{pages}{5188} (\bibinfo{year}{1976}).

\bibitem[{\citenamefont{Perdew and Zunger}(1981)}]{81B-Perdew}
\bibinfo{author}{\bibfnamefont{J.~P.} \bibnamefont{Perdew}} \bibnamefont{and}
  \bibinfo{author}{\bibfnamefont{A.}~\bibnamefont{Zunger}},
  \bibinfo{journal}{Phys. Rev. B} \textbf{\bibinfo{volume}{23}},
  \bibinfo{pages}{5048} (\bibinfo{year}{1981}).

\bibitem[{\citenamefont{Huang and Carter}(2008)}]{08PCCP-Huang-BLPS}
\bibinfo{author}{\bibfnamefont{C.}~\bibnamefont{Huang}} \bibnamefont{and}
  \bibinfo{author}{\bibfnamefont{E.~A.} \bibnamefont{Carter}},
  \bibinfo{journal}{Phys. Chem. Chem. Phys.} \textbf{\bibinfo{volume}{10}},
  \bibinfo{pages}{7109} (\bibinfo{year}{2008}).

\bibitem[{\citenamefont{Murnaghan}(1944)}]{44-Murnaghan-bulkmodulus}
\bibinfo{author}{\bibfnamefont{F.}~\bibnamefont{Murnaghan}},
  \bibinfo{journal}{Proc. Natl. Acad. Sci.} \textbf{\bibinfo{volume}{30}},
  \bibinfo{pages}{244} (\bibinfo{year}{1944}).

\bibitem[{\citenamefont{Ho et~al.}(2008)\citenamefont{Ho, Lign{\`e}res, and
  Carter}}]{08CPC-Ho-profess}
\bibinfo{author}{\bibfnamefont{G.~S.} \bibnamefont{Ho}},
  \bibinfo{author}{\bibfnamefont{V.~L.} \bibnamefont{Lign{\`e}res}},
  \bibnamefont{and} \bibinfo{author}{\bibfnamefont{E.~A.}
  \bibnamefont{Carter}}, \bibinfo{journal}{Comp. Phys. Commun.}
  \textbf{\bibinfo{volume}{179}}, \bibinfo{pages}{839} (\bibinfo{year}{2008}).

\bibitem[{\citenamefont{Gillan}(1989)}]{89JP-Gillan-vacancy}
\bibinfo{author}{\bibfnamefont{M.}~\bibnamefont{Gillan}}, \bibinfo{journal}{J.
  Phys.: Condens. Matter} \textbf{\bibinfo{volume}{1}}, \bibinfo{pages}{689}
  (\bibinfo{year}{1989}).

\bibitem[{\citenamefont{Bernstein and Tadmor}(2004)}]{04B-bernstein-stack}
\bibinfo{author}{\bibfnamefont{N.}~\bibnamefont{Bernstein}} \bibnamefont{and}
  \bibinfo{author}{\bibfnamefont{E.}~\bibnamefont{Tadmor}},
  \bibinfo{journal}{Phys. Rev. B} \textbf{\bibinfo{volume}{69}},
  \bibinfo{pages}{094116} (\bibinfo{year}{2004}).

\bibitem[{\citenamefont{Tallon and
  Wolfenden}(1979)}]{79JPCS-Tallon-al_modulus_exp}
\bibinfo{author}{\bibfnamefont{J.}~\bibnamefont{Tallon}} \bibnamefont{and}
  \bibinfo{author}{\bibfnamefont{A.}~\bibnamefont{Wolfenden}},
  \bibinfo{journal}{J. Phys. Chem. Solids} \textbf{\bibinfo{volume}{40}},
  \bibinfo{pages}{831} (\bibinfo{year}{1979}).

\bibitem[{\citenamefont{Straumanis and
  Woodward}(1971)}]{71ACSA-Straumanis-al_lattice_exp}
\bibinfo{author}{\bibfnamefont{M.}~\bibnamefont{Straumanis}} \bibnamefont{and}
  \bibinfo{author}{\bibfnamefont{C.}~\bibnamefont{Woodward}},
  \bibinfo{journal}{Acta Crystallographica Section A: Crystal Physics,
  Diffraction, Theoretical and General Crystallography}
  \textbf{\bibinfo{volume}{27}}, \bibinfo{pages}{549} (\bibinfo{year}{1971}).

\bibitem[{\citenamefont{Triftsh{\"a}user}(1975)}]{75B-Triftshauser-al_vacancy_exp}
\bibinfo{author}{\bibfnamefont{W.}~\bibnamefont{Triftsh{\"a}user}},
  \bibinfo{journal}{Phys. Rev. B} \textbf{\bibinfo{volume}{12}},
  \bibinfo{pages}{4634} (\bibinfo{year}{1975}).

\bibitem[{\citenamefont{Watkins and
  Corbett}(1964)}]{64PR-Watkins-si_vacancy_exp1}
\bibinfo{author}{\bibfnamefont{G.}~\bibnamefont{Watkins}} \bibnamefont{and}
  \bibinfo{author}{\bibfnamefont{J.}~\bibnamefont{Corbett}},
  \bibinfo{journal}{Phys. Rev.} \textbf{\bibinfo{volume}{134}},
  \bibinfo{pages}{A1359} (\bibinfo{year}{1964}).

\bibitem[{\citenamefont{Dannefaer et~al.}(1986)\citenamefont{Dannefaer,
  Mascher, and Kerr}}]{86L-Dannefaer-si_vacancy_exp2}
\bibinfo{author}{\bibfnamefont{S.}~\bibnamefont{Dannefaer}},
  \bibinfo{author}{\bibfnamefont{P.}~\bibnamefont{Mascher}}, \bibnamefont{and}
  \bibinfo{author}{\bibfnamefont{D.}~\bibnamefont{Kerr}},
  \bibinfo{journal}{Phys. Rev. Lett.} \textbf{\bibinfo{volume}{56}},
  \bibinfo{pages}{2195} (\bibinfo{year}{1986}).

\bibitem[{\citenamefont{Huang and Carter}(2012)}]{12B-Huang-density-decomp}
\bibinfo{author}{\bibfnamefont{C.}~\bibnamefont{Huang}} \bibnamefont{and}
  \bibinfo{author}{\bibfnamefont{E.~A.} \bibnamefont{Carter}},
  \bibinfo{journal}{Phys. Rev. B} \textbf{\bibinfo{volume}{85}},
  \bibinfo{pages}{045126} (\bibinfo{year}{2012}).

\bibitem[{\citenamefont{Ke et~al.}(2013)\citenamefont{Ke, Libisch, Xia, Wang,
  and Carter}}]{13L-Ke-amd}
\bibinfo{author}{\bibfnamefont{Y.}~\bibnamefont{Ke}},
  \bibinfo{author}{\bibfnamefont{F.}~\bibnamefont{Libisch}},
  \bibinfo{author}{\bibfnamefont{J.}~\bibnamefont{Xia}},
  \bibinfo{author}{\bibfnamefont{L.-W.} \bibnamefont{Wang}}, \bibnamefont{and}
  \bibinfo{author}{\bibfnamefont{E.~A.} \bibnamefont{Carter}},
  \bibinfo{journal}{Phys. Rev. Lett.} \textbf{\bibinfo{volume}{111}},
  \bibinfo{pages}{066402} (\bibinfo{year}{2013}).

\bibitem[{\citenamefont{Ke et~al.}(2014)\citenamefont{Ke, Libisch, Xia, and
  Carter}}]{14B-Ke-amd}
\bibinfo{author}{\bibfnamefont{Y.}~\bibnamefont{Ke}},
  \bibinfo{author}{\bibfnamefont{F.}~\bibnamefont{Libisch}},
  \bibinfo{author}{\bibfnamefont{J.}~\bibnamefont{Xia}}, \bibnamefont{and}
  \bibinfo{author}{\bibfnamefont{E.~A.} \bibnamefont{Carter}},
  \bibinfo{journal}{Phys. Rev. B} \textbf{\bibinfo{volume}{89}},
  \bibinfo{pages}{155112} (\bibinfo{year}{2014}).

\bibitem[{\citenamefont{Martienssen and
  Warlimont}(2006)}]{06-Martienssen-si_bulk_exp}
\bibinfo{author}{\bibfnamefont{W.}~\bibnamefont{Martienssen}} \bibnamefont{and}
  \bibinfo{author}{\bibfnamefont{H.}~\bibnamefont{Warlimont}},
  \emph{\bibinfo{title}{Springer handbook of condensed matter and materials
  data}} (\bibinfo{publisher}{Springer Science \& Business Media},
  \bibinfo{year}{2006}).

\bibitem[{\citenamefont{Zhou et~al.}(2004)\citenamefont{Zhou, Wang, and
  Carter}}]{04B-Zhou-pseudo}
\bibinfo{author}{\bibfnamefont{B.}~\bibnamefont{Zhou}},
  \bibinfo{author}{\bibfnamefont{Y.~A.} \bibnamefont{Wang}}, \bibnamefont{and}
  \bibinfo{author}{\bibfnamefont{E.~A.} \bibnamefont{Carter}},
  \bibinfo{journal}{Phys. Rev. B} \textbf{\bibinfo{volume}{69}},
  \bibinfo{pages}{125109} (\bibinfo{year}{2004}).

\bibitem[{\citenamefont{Legrain and Manzhos}(2015)}]{15CPL-Legrain-pesudo}
\bibinfo{author}{\bibfnamefont{F.}~\bibnamefont{Legrain}} \bibnamefont{and}
  \bibinfo{author}{\bibfnamefont{S.}~\bibnamefont{Manzhos}},
  \bibinfo{journal}{Chem. Phys. Lett.} \textbf{\bibinfo{volume}{622}},
  \bibinfo{pages}{99} (\bibinfo{year}{2015}).

\bibitem[{\citenamefont{Del~Rio et~al.}(2017)\citenamefont{Del~Rio, Dieterich,
  and Carter}}]{17JCTC-Del-pseudo}
\bibinfo{author}{\bibfnamefont{B.~G.} \bibnamefont{Del~Rio}},
  \bibinfo{author}{\bibfnamefont{J.~M.} \bibnamefont{Dieterich}},
  \bibnamefont{and} \bibinfo{author}{\bibfnamefont{E.~A.}
  \bibnamefont{Carter}}, \bibinfo{journal}{J. Chem. Theory Comput.}
  \textbf{\bibinfo{volume}{13}}, \bibinfo{pages}{3684} (\bibinfo{year}{2017}).

\end{thebibliography}
\end{document}

% \begin{frontmatter}

\title{Supporting Information: \\ Truncated Non-Local Kinetic Energy Density Functionals for \mc{Simple Metals} and \mc{Silicon}}

\author{Liang Sun}
\affiliation{HEDPS, CAPT, School of Physics and College of Engineering, Peking University, Beijing 100871, P. R. China}
\author{Yuanbo Li}
\affiliation{HEDPS, CAPT, School of Physics and College of Engineering, Peking University, Beijing 100871, P. R. China}
\author{Mohan Chen}
\email{mohanchen@pku.edu.cn}
\affiliation{HEDPS, CAPT, School of Physics and College of Engineering, Peking University, Beijing 100871, P. R. China}
\affiliation{AI for Science Institute, Beijing 100080, P. R. China}
\date{\today}
\pacs{71.15.Mb, 71.20.Mq}
% \end{frontmatter}

\maketitle

\section{Parameters used in OFDFT and KSDFT calculations}

The detailed parameters of KSDFT and OFDFT calculations are shown in Table~\ref{tab:parameters}, including the energy cutoffs and the $k$-point samplings of KSDFT calculations, as well as the energy cutoffs used in OFDFT.

%-------------
% Table S1
%-------------
\begin{table*}[ht]
    \centering
    \caption{Energy cutoff ($E_{\tx{cut}}$ in eV) and $k$-point mesh of KSDFT, and energy cutoff ($E_{\tx{cut}}$ in eV) of OFDFT.}
    \begin{tabularx}{0.9\linewidth}{
    >{\raggedright\arraybackslash\hsize=1.9\hsize\linewidth=\hsize}X
    >{\centering\arraybackslash\hsize=.55\hsize\linewidth=\hsize}X    >{\centering\arraybackslash\hsize=.55\hsize\linewidth=\hsize}X
    >{\centering\arraybackslash\hsize=1\hsize\linewidth=\hsize}X}
    \hline\hline
                                           &OFDFT   &KSDFT  &KSDFT\\
                                           % &OFDFT   &\multicolumn{2}{c}{KSDFT}\\
        System                             &$E_{\tx{cut}}$&$E_{\tx{cut}}$&$k$-point mesh\\\hline
        fcc, bcc, sc Al                    &$800$&$800$&$20 \times 20 \times 20$\\
        hcp Al, hcp Mg                     &$800$&$800$&$12 \times 12 \times 12$\\
        Al fcc(111) surface                &$800$&$800$&$18 \times 18 \times 1$\\
        Al fcc(100), (110) surfaces     &$800$&$800$&$20 \times 20 \times 1$\\
        Al fcc$(1\times1\times1)$ supercell&$800$&$800$&$20 \times 20 \times 20$\\
        Al fcc$(2\times1\times1)$ supercell&$800$&$800$&$10 \times 20 \times 20$\\
        Al fcc$(  2\times2\times1)$ supercell&$800$&$800$&$8 \times 8 \times 16$\\
        Al fcc$(2\times2\times2)$ supercell&$800$&$800$&$8 \times 8 \times 8$\\
        Stacking fault energies in fcc Al (20 layers)  &$800$&$480$&$10 \times 10 \times 1$\\
        Stacking fault energies in fcc Al (22 layers)  &$800$&$480$&$10 \times 10 \times 1$\\
        CD, $\beta$-tin Si                 &$1000$&$800$&$12 \times 12 \times 12$\\
        Si CD(100) surface                 &$1000$&$900$&$12 \times 12 \times 1$\\
        Si CD$(1\times1\times1)$ supercell &$1000$&$800$&$12 \times 12 \times 12$\\
        Si CD$(2\times1\times1)$ supercell &$1000$&$800$&$6 \times 12 \times 12$\\
        Si CD$(2\times2\times2)$ supercell &$1000$&$800$&$8 \times 8 \times 8$\\
        bcc, fcc, sc, CD Li                &$800$&$800$&$20 \times 20 \times 20$\\
        fcc, bcc, sc Mg                    &$800$&$800$&$20 \times 20 \times 20$\\
        \hline\hline
    \end{tabularx}
    \label{tab:parameters}
\end{table*}

\section{Optimization of Truncated Kinetic Kernels}
% \section{Parameters of Truncated Kinetic Kernels}

The truncated kinetic kernels constructed in this work consist of eight Spherical Bessel functions, and the coefficients of these functions are optimized using the simulated annealing method.

The optimized coefficients for \TKK{m}{}s and \TKK{s}{}s are listed in Table~\ref{tab:parameters}.

%-------------
% Table S2
%-------------
\begin{table*}[htbp]
	\centering
	\caption{The coefficients of \TKK{m}{}s and \TKK{s}{}s, $c_i$ is the coefficient of the $i^{\rm{th}}$ Spherical Bessel function in Eq.4.}
	\begin{tabularx}{0.99\linewidth}{
			>{\centering\arraybackslash}X
			>{\raggedleft\arraybackslash}X
			>{\raggedleft\arraybackslash}X
			>{\raggedleft\arraybackslash}X
			>{\raggedleft\arraybackslash}X
			>{\raggedleft\arraybackslash}X
			>{\raggedleft\arraybackslash}X}
		\hline\hline
		Coefficients &{TKK$^{\rm{m}}_{8}$} 	&{TKK$^{\rm{m}}_{12}$}    &{TKK$^{\rm{m}}_{16}$}    &{TKK$^{\rm{s}}_{8}$} &{TKK$^{\rm{s}}_{12}$}    &{TKK$^{\rm{s}}_{16}$} \\
		\hline
$c_1$    & $3.15274$e$-$3   & $1.19734$e$-$3   & $5.48093$e$-$4   & $3.30735$e$-$3   & $1.93888$e$-$3   & $7.61877$e$-$4\\
$c_2$    & $1.65886$e$-$4   & $2.83447$e$-$3   & $1.44417$e$-$3   & $9.39512$e$-$4   & $3.46579$e$-$3   & $2.49824$e$-$3\\
$c_3$    &$-4.91858$e$-$3   & $8.69688$e$-$4   & $2.57833$e$-$3   &$-2.36730$e$-$2   & $1.61192$e$-$3   & $2.16296$e$-$3\\
$c_4$    &$-3.36048$e$-$3   &$-8.03775$e$-$3   &$-3.63401$e$-$7   & $3.76058$e$-$2   &$-1.30013$e$-$2   & $1.32278$e$-$3\\
$c_5$    &$-3.82667$e$-$3   & $3.38437$e$-$3   &$-3.92908$e$-$3   & $2.14557$e$-$2   &$-1.73982$e$-$3   &$-5.87812$e$-$3\\
$c_6$    &$-1.47054$e$-$2   &$-1.87532$e$-$3   &$-4.04051$e$-$3   &$-6.63402$e$-$4   &$-5.63089$e$-$3   &$-1.06169$e$-$2\\
$c_7$    &$-1.38724$e$-$2   &$-1.55350$e$-$2   &$-1.46224$e$-$3   & $1.13403$e$-$1   &$-3.84815$e$-$2   & $2.08889$e$-$3\\
$c_8$    & $1.87566$e$-$2   & $1.47871$e$-$2   & $5.88883$e$-$3   &$-7.51445$e$-$2   & $4.08864$e$-$2   & $5.79804$e$-$3\\
		\hline\hline
	\end{tabularx}
	\label{tab:Coef}
\end{table*}

\begin{figure}[hbp]
    \centering
    
    \begin{subfigure}{0.48\textwidth}
    \centering
    \includegraphics[width=0.98\linewidth]{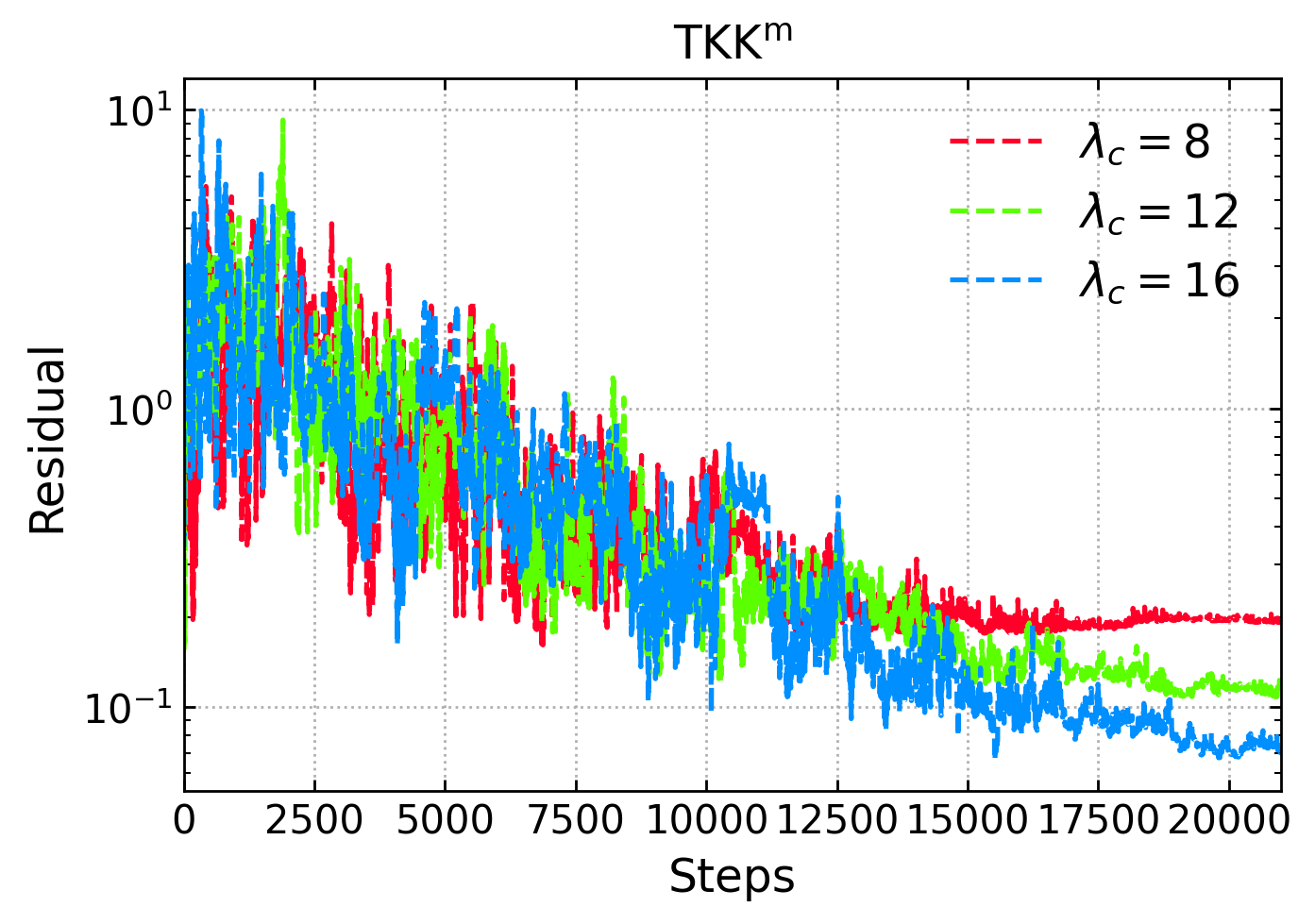}
    \end{subfigure}
    \begin{subfigure}{0.48\textwidth}
    \centering
    \includegraphics[width=0.98\linewidth]{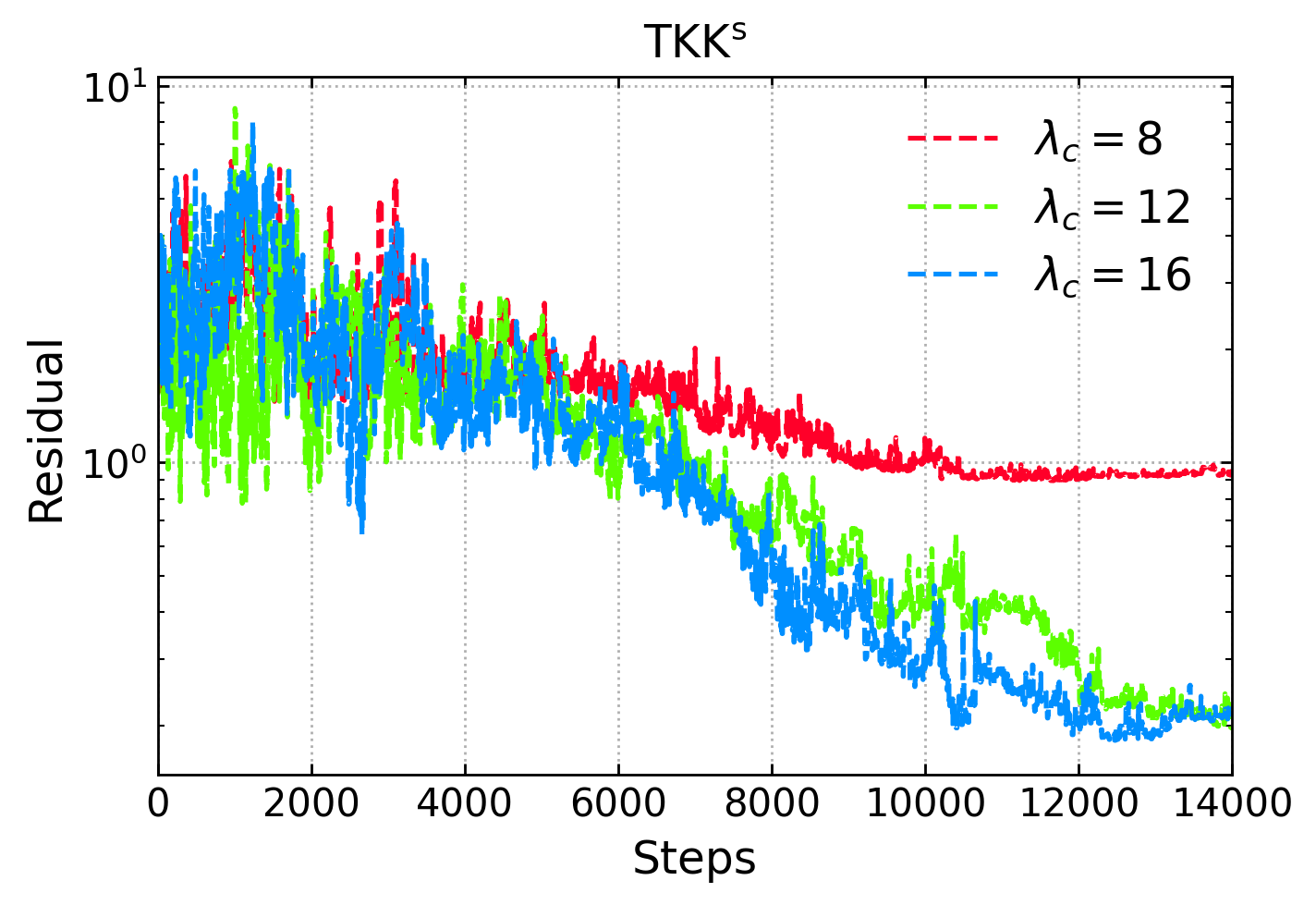}
    \end{subfigure}
    
    \caption{\mc{Changes of residual (defined in Eq.~5) with respect to the optimization step for a few (a) \TKK{m}{}s and (b) \TKK{s}{}s.}}
    \label{fig:SI_residual}
\end{figure}

\mc{The residuals (defined in Eq.~5) of both \TKK{m}{}s and \TKK{s}{}s with different cutoffs during the optimization are displayed in Fig.~\ref{fig:SI_residual}.}

%-------------
% Figure S1
%-------------

\section{Stacking Fault Energies}

\begin{figure*}[htbp]
    \centering
    
    \begin{subfigure}{0.48\textwidth}
    \centering
    \includegraphics[width=0.98\linewidth]{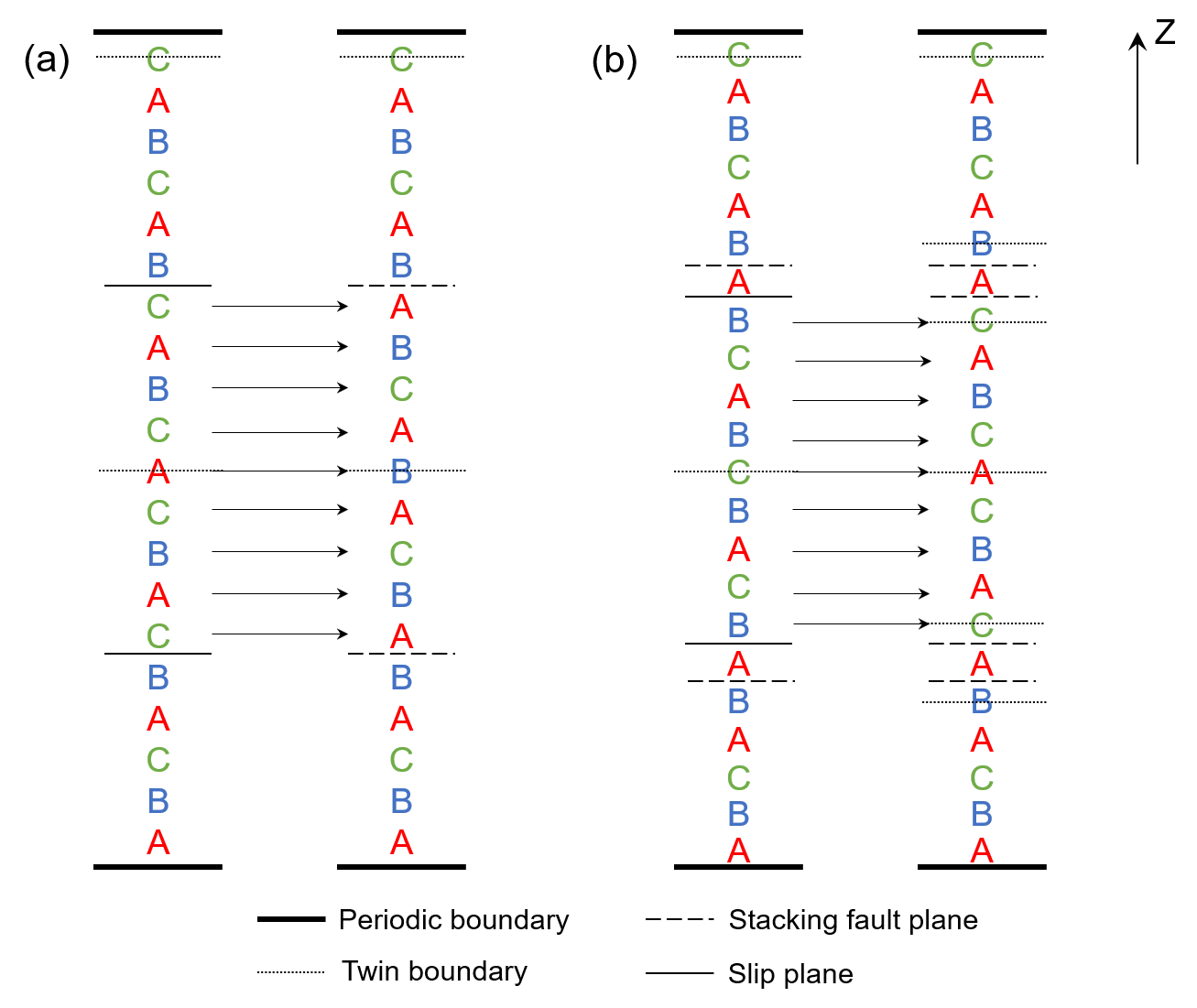}
    \label{fig:SI_Stack_Alab}
    \end{subfigure}
    \begin{subfigure}{0.50\textwidth}
    \centering
    \includegraphics[width=0.98\linewidth]{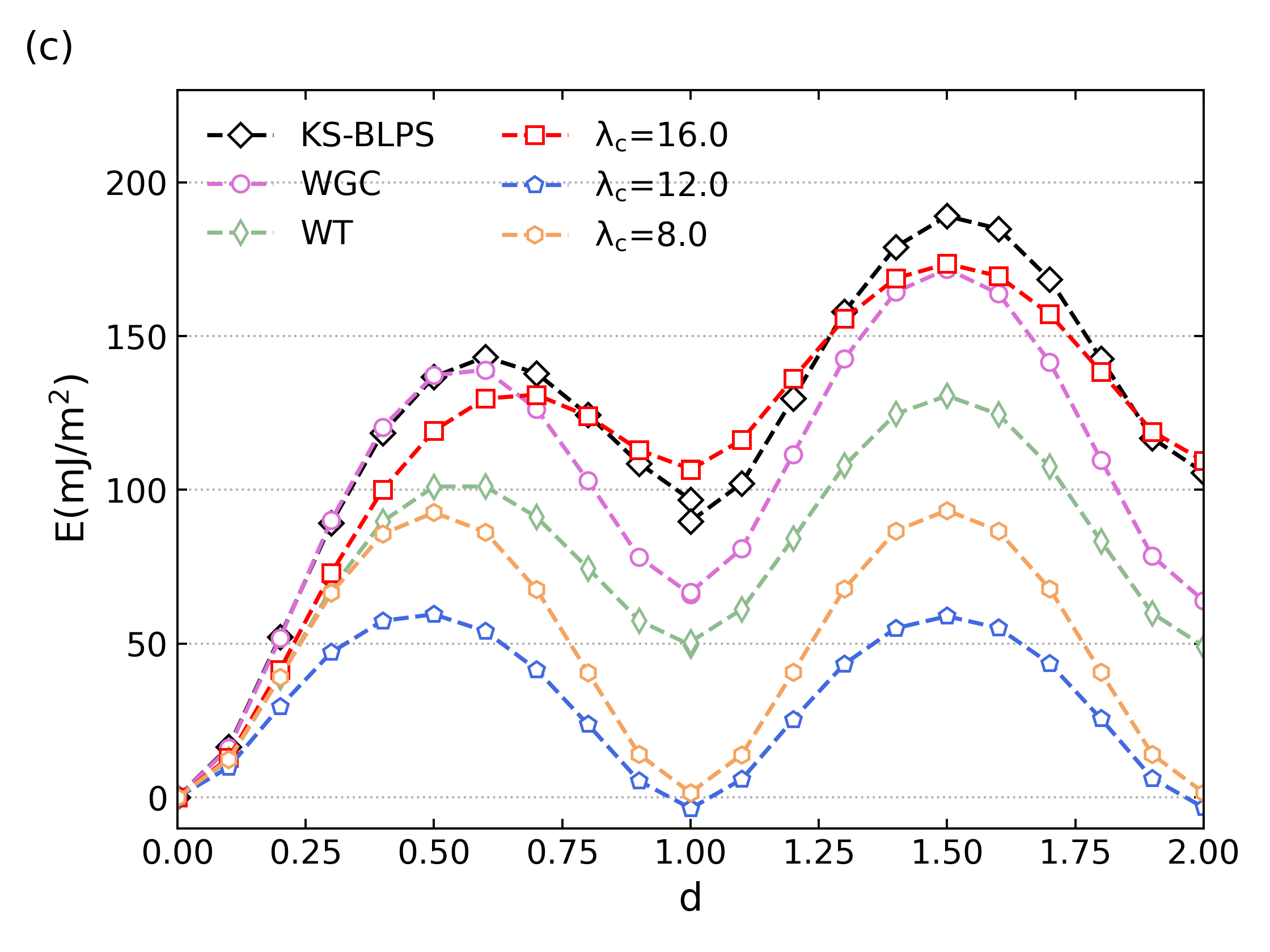}
    \label{fig:SI_Stack_Alc}
    \end{subfigure}
    
    \caption{Figures (a) and (b) illustrate the schematic diagram (after Bernstein and Tadmor\cite{04B-bernstein-stack}) of the stacking fault in fcc Al. Here A, B, and C denote three kinds of atomic layers in the fcc Al structure, Z depicts the $[11\bar{1}]$ direction, and the arrows indicate the lateral translation of atomic layers. 
    Figures (a) and (b) illustrate the deformation processes from $d = 0.0$ to $1.0$ and from $d = 1.0$ to $2.0$ in (c), respectively. 
    Figure (c) shows the computed stacking fault energies of fcc Al ($E$, in mJ $\tx{m}^{-2}$) as a function of the total fractional displacement $d$, which is the translation along the [112] direction with ten steps. The step interval is $\frac{1}{10}\frac{a}{\sqrt{6}}$ with $a$ being the equilibrium lattice constant of fcc Al. 
    Only twin boundaries exist at $d=0.0$. The unstable stacking fault forms at $d\approx0.5$, the intrinsic stacking fault forms at $d\approx1.0$, the unstable twinning fault forms at $d\approx1.5$, and the extrinsic stacking fault forms at $d\approx2.0$. We use both OFDFT and KSDFT calculations. A variety of KEDFs are used in OFDFT calculations and explained in the main text. The results of KSDFT are obtained from Ref.~\onlinecite{08PCCP-Huang-BLPS}.}
    \label{fig:SI_Stack_Al}
\end{figure*}

The stacking fault configurations in fcc Al as well as the corresponding computational results are displayed in Fig.~\ref{fig:SI_Stack_Al}.
As shown in Fig.~S2(a), (b), “A, B, C” denotes three kinds of atomic layers in fcc Al, and the arrows indicate the lateral translation of atomic layers.
Each atomic layer contains one atom.
To obtain stacking fault energies, after each small translation, we performed structure optimization in the z-direction, with lattice vectors in the x-y plane fixed.
The energy changes during the translation process were recorded and are presented in Fig.~S2(c).

% -------------
% Figure S2
% -------------

\section{Mg-Al Alloys}

\mc{In order to further assess the transferability of the \TKK{m}{16} KEDF, we have performed KSDFT and OFDFT calculations for two Mg-Al alloys, i.e., $\beta''\ \tx{Al}_3\tx{Mg}$~\cite{03MSMSE-Carling} and $\tx{L1}_2\ \tx{Mg}_3\tx{Al}$~\cite{17MSMSE-Zhuang}.
The bulk moduli and equilibrium volumes of the two alloys are listed in Table.~\ref{tab:MgAl_bulk}. As shown in the table, we find the WGC, WT, and \TKK{m}{16} KEDFs are able to reproduce the bulk moduli and equilibrium volumes of the two Mg-Al alloys obtained by KSDFT.}
%
% However, the formation energies obtained by various methods are quite different from each other, and none of three KEDFs can reproduce the results got obtained by KSDFT.}

% \SL{This issue may be attribute to two reasons: 1. The formation energies, in meV, are quite small, and are too sensitive to the method, so that is hard to reproduce. 2. There are no alloys in the training set. We note that the results obtained by TKKm16 KEDF are not worse than those got by WT KEDF.}

%-------------------------------
% Table S3: Mg-Al alloys
%-------------------------------
\begin{table*}[htbp]
	\centering
	% \caption{\SL{OFDFT and KSDFT results for bulk modulus ($B$ in $\tx{GPa}$), equilibrium volume ($V_0$ in $\tx{\AA}^3$ per atom), and formation energy (${\Delta}E_{\tx{f}}$ in meV) of two kinds of Mg-Al alloys.}}
 	\caption{\mc{OFDFT and KSDFT results for the bulk moduli ($B$ in $\tx{GPa}$) and equilibrium volumes ($V_0$ in $\tx{\AA}^3$/atom) of two Mg-Al alloys.
    Both KSDFT and OFDFT calculations are performed with the use of bulk-derived pseudopotentials (BLPS). 
    For OFDFT calculations, we use the WGC and WT KEDFs.}}
	\begin{tabularx}{0.75\linewidth}{
			>{\raggedright\arraybackslash}X
			>{\raggedright\arraybackslash}X
			>{\centering\arraybackslash}X
			>{\centering\arraybackslash}X}
		\hline\hline
		Mg-Al alloys    && $\beta''\ \tx{Al}_3\tx{Mg}$    &$\tx{L1}_2\ \tx{Mg}_3\tx{Al}$\\
		\hline
		$B\ (\tx{GPa})$         &KS-BLPS          &67  &48  \\
		% &truncated WT   &&62.2  &&39.3  &\\
		&WGC            &67  &47  \\
		&WT             &63  &39  \\
            &\TKK{m}{16}   &64  &39  \\
		\\
		$V_0\ (\tx{\AA}^3)$ &KS-BLPS          &16.773  &19.547  \\
		% &truncated WT   &&17.450  &&20.645  &\\
		&WGC            &16.786  &19.469  \\
		&WT             &17.152  &20.270  \\
            &\TKK{m}{16}   &16.956  &20.138  \\
		% \\
		% ${\Delta}E_{\tx{f}}\ (\tx{meV})$  &\TKK{m}{16}   &11.4      &$-25.0$   \\
		% % &truncated WT   &&27.4      &&$-27.1$   &\\
		% &WT             &32.8      &$-19.1$   \\
		% &WGC            &$-6.2$    &5.8       \\
		% &KS-BLPS          &$-9.0$    &$-15.7$   \\
		\hline\hline
	\end{tabularx}
	\label{tab:MgAl_bulk}
\end{table*}

\section{HD and cbcc Phases of Si}

\mc{To test the transferability of the \TKK{s}{16} KEDF, we have performed KSDFT and OFDFT calculations for the hexagonal diamond (HD) and complex body-centered-cubic (cbcc) phases of Si, and the results are listed in Table.~\ref{tab:HD_cbcc_Si}. For OFDFT calculations, we adopted both HC~\cite{10B-Huang-nonlocal} and \TKK{s}{16} KEDFs. In general, we observe that the \TKK{s}{16} KEDF yields the correct energy ordering of the CD, HD, and cbcc phases; however, the energy differences predicted by the \TKK{s}{16} KEDF are slightly larger than those obtained by the HC and the KS-BLPS methods. Additionally, the bulk moduli obtained by the \TKK{s}{16} KEDF are smaller than the other two methods. In conclusion, the accuracy of the \TKK{s}{16} KEDF is slightly lower than the HC KEDF, which is expected because only the CD data are used to optimize the \TKK{s}{16} KEDF kernel. In future, additional phases of Si and other semiconductor systems can be included to further optimize the KEDF kernels.
}

%-------------------------------
% Table S4: HD and cbcc Si
%-------------------------------
\begin{table*}[htbp]
	\centering
	\caption{\mc{OFDFT and KSDFT results for the bulk moduli ($B$ in $\tx{GPa}$), equilibrium volumes ($V_0$ in $\tx{\AA}^3$/atom), and ground state energies ($E_0$ in eV/atom) of the hexagonal diamond (HD) and complex body-centered-cubic (cbcc) phases of Si. 
    %
    %We set the $E_0$ to be the total energy for the CD structure, 
    The total energy of the HD/cbcc phase is set to the energy difference with respect to the total energy of the CD phase of Si.
    Both KSDFT and OFDFT calculations are performed with the usage of BLPS. 
    For OFDFT calculations, the HC KEDF~\cite{10B-Huang-nonlocal} and the \TKK{s}{16} KEDF are utilized.}}
	\begin{tabularx}{0.75\linewidth}{
			>{\raggedright\arraybackslash}X
			>{\raggedright\arraybackslash}X
			>{\centering\arraybackslash}X
			>{\centering\arraybackslash}X
			>{\centering\arraybackslash}X}
		\hline\hline
		    && CD    &HD  &cbcc\\
		\hline
		$B\ (\tx{GPa})$         &KS-BLPS~\cite{10B-Huang-nonlocal}    &99  &99    &101  \\
		&HC~\cite{10B-Huang-nonlocal} &97  &98  &105\\
            &\TKK{s}{16}   &78  &81  &77\\
		\\
		$V_0\ (\tx{\AA}^3)$ &KS-BLPS~\cite{10B-Huang-nonlocal}    &19.774  &19.641  &17.594\\
		&HC~\cite{10B-Huang-nonlocal} &19.962  &19.875  &18.419\\
            &\TKK{s}{16}   &19.470  &19.159  &17.385\\
		\\
		$E_0\ (eV)$ &KS-BLPS~\cite{10B-Huang-nonlocal}    &$-109.629$  &0.014  &0.157\\
		&HC~\cite{10B-Huang-nonlocal} &$-109.624$  &0.007  &0.141\\
            &\TKK{s}{16}   &$-109.583$  &0.184  &0.227\\
		\hline\hline
	\end{tabularx}
	\label{tab:HD_cbcc_Si}
\end{table*}

\section{Vacancy Formation Energies}

\begin{figure}[tbp]
    \centering
    
    \begin{subfigure}{0.48\textwidth}
    \centering
    \includegraphics[width=0.98\linewidth]{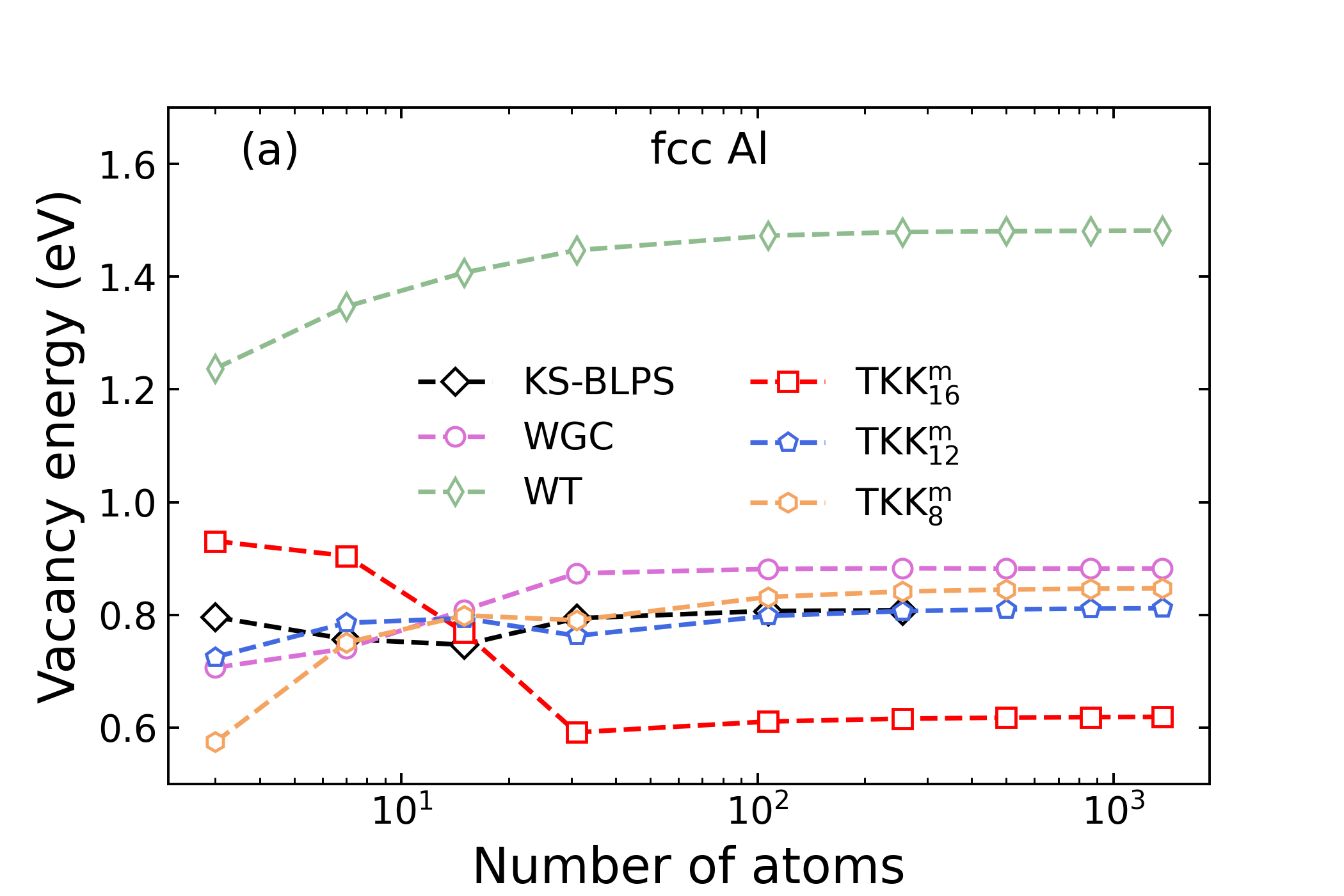}
    % \caption{\label{fig:Vacancy_al}}
    \end{subfigure}
    \begin{subfigure}{0.48\textwidth}
    \centering
    \includegraphics[width=0.98\linewidth]{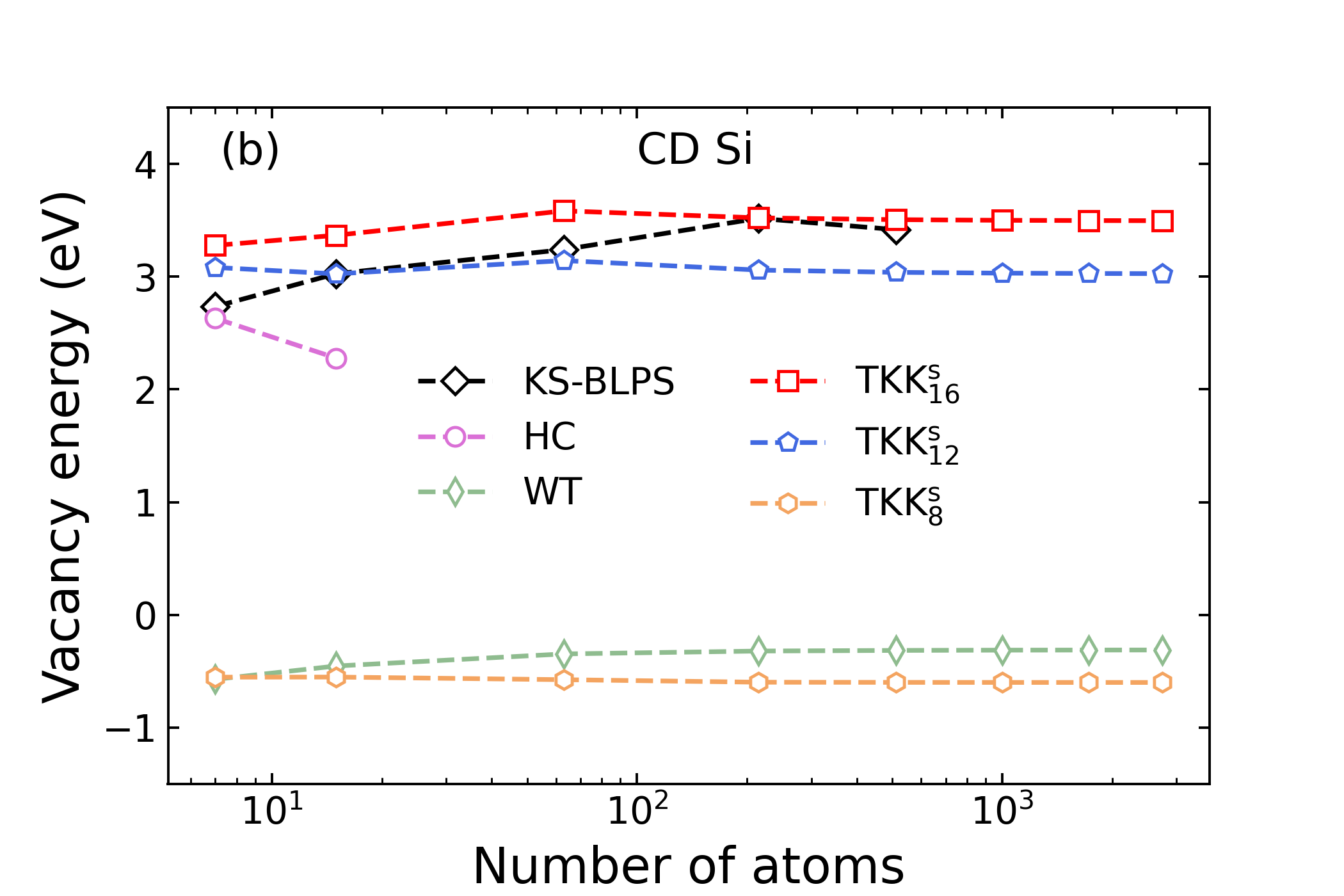}
    % \caption{}
    \end{subfigure}
    
    \caption{Vacancy energies (in eV) of (a) fcc Al and (b) CD Si as computed from KS-BLPS and OFDFT methods with various kinds of KEDFs. The system sizes are 3, 7, 15, 31, 107, 255, 499, 863, 1371 atoms for fcc Al, and 7, 15, 63, 215, 511, 999, 1727, 2743 atoms for CD Si.}
    \label{fig:Vacancy}
\end{figure}

Fig.~\ref{fig:Vacancy} shows the vacancy formation energies of fcc Al and CD Si as computed by both OFDFT and KSDFT calculations. The system size is chosen from 3 to 1371 atoms in fcc Al and ranges from 7 to 2743 atoms for CD Si.

%-------------
% Figure S3
%-------------

\section{Surface Configurations}

\begin{figure}[tbp]
    \centering
    
    \begin{subfigure}{0.6\textwidth}
    \centering
    \includegraphics[width=0.98\linewidth]{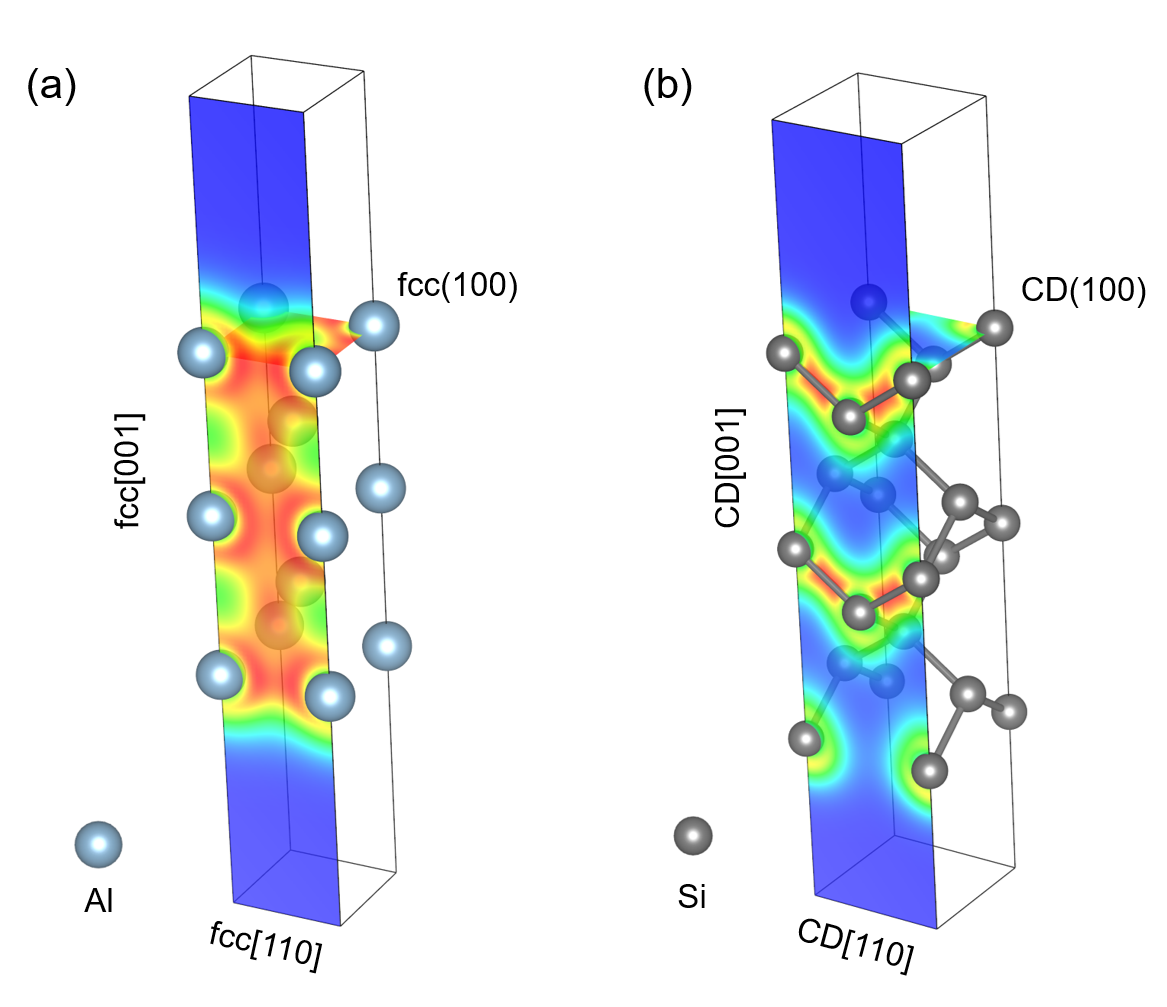}
    \label{fig:slab}
    \end{subfigure}
    
    \caption{The configurations of (a) Al fcc(100) surface and (b) Si CD(100) surface with the electron densities on the longitudinal section. The calculations are obtained from KSDFT.}
    \label{fig:Stack_Al}
\end{figure}

Figs.~\ref{fig:Stack_Al}(a) and (b) illustrate the surface configurations of Al fcc (100) and Si CD (100), respectively.

%-------------
% Figure S4
%-------------

\bibliography{OF-KEDF}